%% file: ms.tex
\documentclass[]{jfm}
\pdfoutput=1
\usepackage{natbib}
\usepackage{amsmath}
\usepackage[dvips]{graphicx}
\usepackage{color}
\usepackage{tikz}

\newcommand{\aver}[1]{\! \left\langle {#1} \right\rangle \!}
\newcommand{\vaver}[1]{\! \left[\, {#1}\, \right] \!}
\newcommand{\ud}{{\rm{d}}}

\newcommand{\Reyn}{Re_\Pi}
\newcommand{\produ}{\mathcal{P}}

\newcommand{\dissU}{\phi}
\newcommand{\dissLam}{\phi_\ell}
\newcommand{\dissDelta}{\phi_\Delta}

\newcommand{\power}{\Pi}
\newcommand{\uv}{r(y)}
\newcommand{\Ulam}{U_\ell}
\newcommand{\UDelta}{U_\Delta}

\begin{document}

\title[]{Global energy fluxes \\ in fully-developed turbulent channels \\ with flow control}

\author[D.Gatti, A.Cimarelli, Y.Hasegawa, B.Frohnapfel \& M.Quadrio]{
D\ls A\ls V\ls I\ls D\ls E\ns G\ls A\ls T\ls T\ls I$^1$ \thanks{Email address for correspondence: davide.gatti@kit.edu}, \ls
A\ls N\ls D\ls R\ls E\ls A\ns C\ls I\ls M\ls A\ls R\ls E\ls L\ls L\ls I$^2$, \ls
Y\ls O\ls S\ls U\ls K\ls E\ns H\ls A\ls S\ls E\ls G\ls A\ls W\ls A$^3$, \ls
B\ls E\ls T\ls T\ls I\ls N\ls A\ns F\ls R\ls O\ls H\ls N\ls A\ls P\ls F\ls E\ls L$^1$
\and \ns
M\ls A\ls U\ls R\ls I\ls Z\ls I\ls O\ls \ns Q\ls U\ls A\ls D\ls R\ls I\ls O$^4$
\thanks{Mercator Fellow at Karlsruhe Institute of Technology}
}

\affiliation{$^1$Institute of Fluid Mechanics, Karlsruhe Institute of Technology,
Kaiserstra\ss e 10, 76131 Karlsruhe, Germany
\\[\affilskip]
$^2$Dipartimento di Ingegneria Industriale e Scienze Matematiche, Universit\`a Politecnica delle Marche, Via Brecce Bianche 12, 60131 Ancona, Italy
\\[\affilskip]
$^3$Institute of Industrial Science, The University of Tokyo, 4-6-1 Komaba, Meguro-ku, Tokyo 153-8505,
Japan
\\[\affilskip]
$^4$Department of Aerospace Science and Technologies, Politecnico di Milano,
via La Masa 34, 20156 Milano, Italy
}

\date{\today}

\maketitle

\begin{abstract}
This paper addresses the integral energy fluxes in natural and controlled turbulent channel flows, where active skin-friction drag reduction techniques allow a more efficient use of the available power. We study whether the increased efficiency shows any general trend in how energy is dissipated by the mean velocity field (mean dissipation) and by the fluctuating velocity field (turbulent dissipation).

Direct Numerical Simulations (DNS) of different control strategies are performed at Constant Power Input (CPI), so that at statistical equilibrium each flow (either uncontrolled or controlled by different means) has the same power input, hence the same global energy flux and, by definition, the same total energy dissipation rate. The simulations reveal that changes in mean and turbulent energy dissipation rates can be of either sign in a successfully controlled flow. 

A quantitative description of these changes is made possible by a new decomposition of the total dissipation, stemming from an extended Reynolds decomposition, where the mean velocity is split into a laminar component and a deviation from it. Thanks to the analytical expressions of the laminar quantities, exact relationships are derived that link the achieved flow rate increase and all energy fluxes in the flow system with 
two wall-normal integrals of the Reynolds shear stress and the Reynolds number. The dependence of the energy fluxes on the Reynolds number is elucidated with a simple model in which the control-dependent changes of the Reynolds shear stress are accounted for via a modification of the mean velocity profile. The physical meaning of the energy fluxes stemming from the new decomposition unveils their inter-relations and connection to flow control, so that a clear target for flow control can be identified.
\end{abstract}

\input{intro}

\input{description}
\input{boxes}
\input{decomposition}
\input{newboxes}
\input{conclusion}

\section*{Acknowledgments}
Support through the Deutsche Forschungsgemeinschaft (DFG) project FR2823/5-1 is gratefully acknowledged. Computing time has been provided by the computational resource ForHLR Phase I funded by the Ministry of Science, Research and the Arts, Baden-W\"urttemberg and DFG. DG gratefully acknowledges the additional support of the Japan Society for the Promotion of Science. YH is supported by the Ministry of Education, Culture, Sports, Science and Technology of Japan (MEXT) through the Grant-in-Aid for Scientific Research (B) (No. 17H03170).

\input{newboxes-Re}

\end{document}

%% file: intro.tex
\section{Introduction}

Drag reduction in turbulent flows, and in particular skin-friction drag reduction in wall-bounded turbulent flows, is attracting scholars since the early days of fluid mechanics. Its goal is achieving a more energetically efficient flow system, yielding large economical savings and emission reductions for the global transport sector. Nowadays, a number of techniques exist, either passive or active and at various levels of technical readiness, which promise to successfully reduce turbulent friction drag. Its successful reduction is not only entailed with practical significance, but also helpful to advance our understanding of the physics of near-wall turbulence. The main question addressed in this paper is: How are the global energy fluxes (i.e. power input, transfer rates between mean and turbulent kinetic energy, and energy dissipation rates) modified in a fluid system where drag reduction enables increased energetic efficiency?

From an energy standpoint, an incompressible turbulent flow, however complex, can be thought of as a dissipative system that requires external energy to operate: homogeneous isotropic turbulence is not sustained without some external energy input, and a fluid in a duct only flows when a pump is present or a pressure gradient is established by means of external forces.
The simplest wall-bounded flow is the plane Poiseuille flow, contained between two parallel, indefinitely large walls. The flux of energy through this system is described, at the highest level of compactness, by the time-averaged rate at which energy enters the system (i.e. the pumping power $\power_p$ and possibly an extra control power $\power_c$ if active flow control is considered), which must equal, at statistical equilibrium, the time-averaged rate at which energy is dissipated via viscous mechanisms (i.e. the total viscous dissipation).

If one adopts the conventional Reynolds decomposition of the velocity vector into a time-averaged and a fluctuating field, pumping energy is associated with the mean flow alone, while energy dissipation rate is split into one part $\dissU$ associated with the dissipation of the time-averaged velocity field, and another part $\epsilon$ associated with the fluctuating field. Hence:
\begin{equation*}
\power_p + \power_c = \dissU + \epsilon .
\end{equation*}

When applied to the kinetic energy of the flow, the Reynolds decomposition 
separates the kinetic energy of the mean flow (MKE) and the turbulent kinetic energy (TKE). Since these two energies are decoupled, a compact energetic description of a channel flow is obtained by means of the so-called energy box \citep{quadrio-2011, ricco-etal-2012}. Though simplified by the volume- and time-averaging, the energy-box description still highlights the energy transfer process from the mean to the fluctuating field, embodied by the production $\produ$ of turbulent kinetic energy (acting as a sink for MKE, but as a source for TKE).

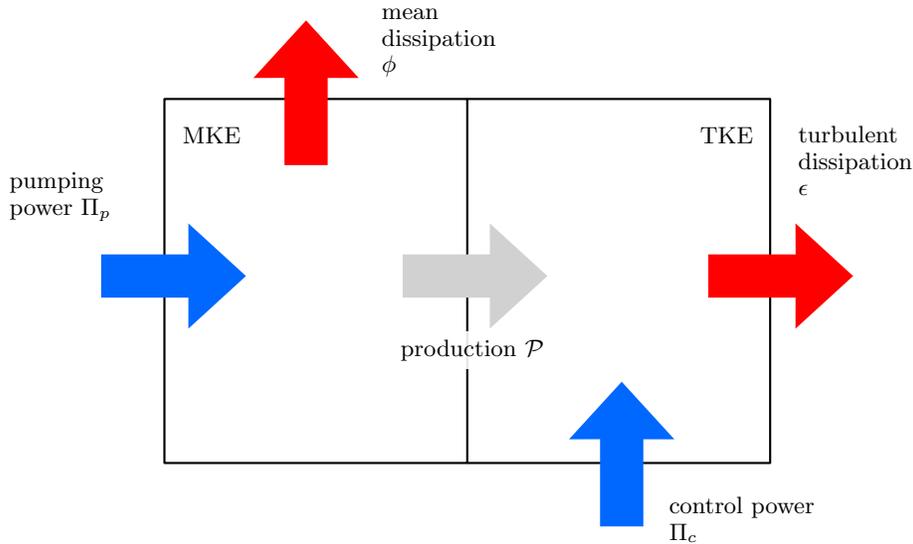
\begin{figure}
\begin{tikzpicture}
\centering
    \node[anchor=south west,inner sep=0] (image) at (0,0) {\includegraphics[]{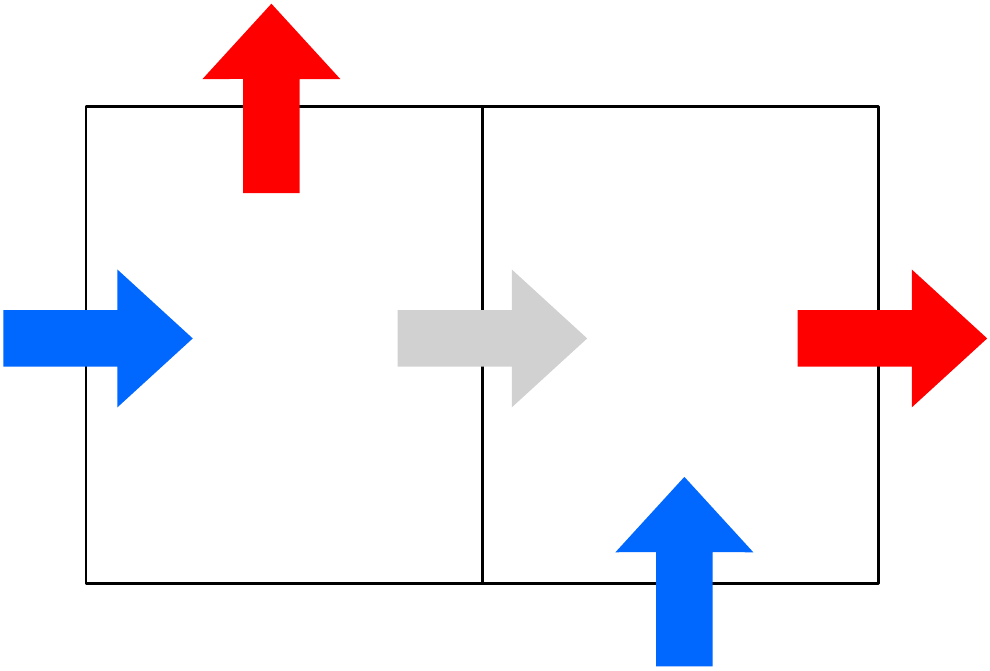}};
    \begin{scope}[x={(image.south east)},y={(image.north west)}]
        \node[align=left] at (0.15,0.77)  {MKE};
        \node[align=left] at (0.83,0.77)  {TKE};
        \node[align=left] at (-0.05,0.65) {pumping\\ power $\Pi_p$};
        \node[align=left] at (0.45,0.95)  {mean\\ dissipation\\ $\dissU$};
        \node[align=left,fill=white,opacity=0.9,text opacity=1] at (0.494,0.35)  {production $\produ$};
        \node[align=left] at (1,0.72)  {turbulent\\ dissipation\\ $\epsilon$};
        \node[align=left] at (0.85,0.02)  {control power\\ $\Pi_c$};
    \end{scope}
\end{tikzpicture}
\caption{Schematic of the energy box for a turbulent channel flow, divided in MKE (mean kinetic energy) and TKE (turbulent kinetic energy) sub-boxes.}
\label{fig:ebox}
\end{figure}

The energy box for a channel flow with control is schematically depicted in Fig.~\ref{fig:ebox}. The control system is assumed to introduce spatial or temporal velocity fluctuations, thus supplying energy to TKE ($\power_c$ in Fig.~\ref{fig:ebox} enters the TKE box). Of course, control strategies do exist \citep[see for instance][]{sumitani-kasagi-1995,xu-etal-2007} which directly supply energy to the mean field; the present approach could be easily modified to account for it. 
Under the assumption that $\power_c$ feeds TKE it holds that:
\begin{equation*}
\power_p =  \phi + \produ, \qquad \power_c + \produ = \epsilon .
\end{equation*}

We want to understand whether an improvement of energetic efficiency via flow control is connected to generally valid changes in one of the energy-dissipating quantities, e.g. an increase/decrease of $\epsilon$. If this is the case, maximising/minimising $\epsilon$ could become the control objective in smart control loops that aim at energetic savings through skin-friction drag reduction; and a substantial hint would be available for the development of RANS turbulence models capable of indirectly representing drag-reduction techniques. In literature, a number of studies discuss the behaviour of $\epsilon$ in drag reduced flows. Some report a decrease of $\epsilon$ \citep{dimitropoulos-etal-2001, ge-etal-2013, agostini-touber-leschziner-2014}, while others an increase of $\epsilon$ \citep{ricco-etal-2012, deangelis-etal-2005}.

The observed variations of $\epsilon$ and $\dissU$ are strongly linked to the way the comparison between the uncontrolled and controlled flow is carried out \citep{frohnapfel-hasegawa-quadrio-2012}. If the two flows are compared by enforcing the same mean pressure gradient (CPG), flow control results in an increase of the flow rate; as a result, the controlled flow is driven with a larger pumping power input, and $\dissU + \epsilon$ increases. In contrast, if one compares two flows at constant flow rate (CFR), the effect of flow control on $\dissU + \epsilon$ is undefined. In fact, drag reduction in the controlled flow yields a decrease of the pressure gradient and thus a decrease of $\power_p$ (given by the product of pressure gradient and flow rate). $\dissU + \epsilon$ may increase, decrease or not vary at all depending on the magnitude of $\power_c$.

The Constant Power Input (CPI) condition has been recently proposed \citep{hasegawa-quadrio-frohnapfel-2014, quadrio-frohnapfel-hasegawa-2016} as an alternative approach for flow control. With CPI, the simulation is set up in such a way that the total power input to the flow system is kept constant. A flow control strategy successfully improves the energetic efficiency of the system when the bulk mean velocity is increased compared to a natural channel flow at the same given total power. Since active flow control requires a certain (often non-negligible) amount of power, the total power $\power_t = \Pi_p + \Pi_c$ is kept constant across different cases, and therefore  $\dissU + \epsilon$ remains constant. This scenario therefore allows studying changes in $\epsilon$ (or $\dissU$) in a more natural way. 

CPI is adopted in the present paper to understand how the energy fluxes in a turbulent channel flow are modified when an increase in energetic efficiency is realised through skin-friction drag reduction. It will be shown that, even in the well-defined CPI setting, changes of $\epsilon$ and $\dissU$ cannot be related to improvements in energetic efficiency. Indeed, $\epsilon$ and $\dissU$ will be expressed in terms of two integrals of the wall-normal Reynolds shear stress distribution and the Reynolds number. Therefore, changes of $\epsilon$ and $\dissU$ are fully determined by the modification in the Reynolds shear stress distribution, in combination with the fraction of power spent for control. We will describe an alternative decomposition of the total viscous dissipation which facilitates the derivation of new analytical results and the physical interpretation of flow control effects onto global energy fluxes and dissipations.

%% file: description.tex
\section{Problem description}

We consider the incompressible, fully-developed flow of a viscous fluid with density $\rho^\ast$, dynamic viscosity $\mu^\ast$ and kinematic viscosity $\nu^\ast$ between two plane parallel walls located $2h^\ast$ apart. $x^\ast$, $y^\ast$ and $z^\ast$ denote the streamwise, wall-normal and spanwise coordinates. The corresponding components of the velocity vector are $u^\ast$, $v^\ast$ and $w^\ast$, and the static pressure is $p^\ast$. Throughout the paper, all dimensional quantities are indicated with an asterisk. Nondimensionalization in viscous (inner) units, i.e. by the kinematic viscosity $\nu^\ast$ and the friction velocity $u_\tau^\ast = \sqrt{\tau_w^\ast / \rho^\ast}$ based upon the wall shear stress $\tau_w^\ast$ of the reference flow, is denoted  by the superscript +.
In all other instances quantities are nondimensionalized by the channel half-height $h^\ast$ and the characteristic velocity $U_\Pi^\ast$, used in the definition of the Reynolds number $\Reyn$, employed in CPI as presented in the following subsection. Angular brackets $ \langle \cdot \rangle $ indicate temporal and spatial averaging across the statistically homogeneous directions. The Reynolds decomposition for a generic quantity $f$ is readily defined as $f = \aver{f} + f'$. The average across the whole fluid domain and in time is denoted by square brackets as $\vaver{f}$.

\subsection{The CPI approach}
In the present work we adopt the constant power input (CPI) framework introduced by \cite{hasegawa-quadrio-frohnapfel-2014}: the flow is driven through the channel by a constant total power $\power_t^\ast$.
The relative importance of the control power $\power_c^\ast$ over the pumping power $\power_p^\ast$ is expressed by the quantity:
\begin{equation*}
  \gamma = \frac{\power_c^\ast}{\power_t^\ast} = 1-\frac{\power_p^\ast}{\power_t^\ast} \ ,
\end{equation*}
which represents the fraction of the total power $\power^*_t$ used by the control, and is therefore zero for the canonical reference flow.

While the control power $\power_c^\ast$ necessarily depends upon the control strategy of choice, the pumping power $\power_p^\ast$ per unit wetted area is given by:
\begin{equation}
  \power^\ast_p = G^\ast h^\ast U_b^\ast
\label{eq:dim-pp}
\end{equation}
where $- G^\ast$ is the mean streamwise pressure gradient, and $U_b^\ast \equiv \vaver{u^\ast}$ is the bulk velocity.

\cite{hasegawa-quadrio-frohnapfel-2014} introduce the velocity $U_\Pi^\ast = \sqrt{\power_t^\ast h^\ast / 3\mu^\ast}$, i.e. the bulk velocity of a laminar flow driven by the pumping pumper $\power_t^\ast$, as the appropriate characteristic velocity in the CPI approach.
This choice is justified by the theoretical argument \citep{bewley-2009,fukagata-sugiyama-kasagi-2009} that a laminar flow maximises the bulk velocity for a given $\power_t^\ast$. The ultimate goal of flow control under CPI is to increase the ratio $U_b^\ast / U_\Pi^\ast$ towards the theoretical upper limit $U_b^\ast / U_\Pi^\ast=1$.

At CPI the corresponding power-based Reynolds number is given by:
\begin{equation*}
  \Reyn = \frac{U^\ast_\Pi h^\ast}{\nu^\ast}
\end{equation*}
which leads directly to the dimensionless power input per unit wetted area:
\begin{equation}
\power_t = \frac{\power_t^\ast}{\rho^\ast {U_\Pi^\ast}^3} = \frac{3}{\Reyn} .
\label{eq:total-power}
\end{equation}
Hence, in CPI a constant $\power_t$  implies working at a constant $\Reyn$, just like CPG corresponds to constant $Re_\tau=u_\tau^\ast h / \nu$ and CFR to a constant $Re_b=U_b^\ast h / \nu$. In case of active control, the available pumping power is:
\begin{equation}
\power_p = \frac{3(1-\gamma)}{\Reyn}.
\label{eq:power-p}
\end{equation}
$\Reyn$, $Re_\tau$ and $Re_b$ are connected through the dimensionless form of Eq.~\eqref{eq:dim-pp}:
\begin{equation}
  3 \left( 1 - \gamma \right) \Reyn^2 = Re_\tau^2 Re_b \,,
  \label{eq:CPIconstraint}
\end{equation}
which highlights once more that $\Reyn$ represents the total power and combines viscous and bulk velocity scales.

\subsection{Numerical details}

The following discussion relies upon analytic as well as numerical results. For the latter, three DNS of turbulent channels have been produced on purpose under the CPI condition. The value of $\Reyn$, kept constant across all cases, is $\Reyn=6500$, corresponding in the uncontrolled (reference) case to $Re_\tau=u_\tau h / \nu=199.7$ and $Re_b= U_b h / \nu = 3176.8$. It is worth recalling that the different forcing strategies CFR, CPG and CPI lead to essentially identical turbulence statistics for wall friction, despite the fact that they produce different dynamical systems, i.e. the temporal behaviour of the space-mean (instantaneous) streamwise velocity and the space-mean (instantaneous) pressure gradient are different \citep{quadrio-frohnapfel-hasegawa-2016}. As discussed before, the choice of the forcing strategy becomes important when two (or more) flows are to be compared with each other. Since the present work is concerned with changes in energy fluxes between a "natural" turbulent flow and drag reduced ones, we choose to compare flows at CPI, such that they all possess the same global energy flux. 

The employed DNS solver is that by \cite{luchini-quadrio-2006}, which uses mixed spatial discretisation (Fourier series in the homogeneous directions, and compact, fourth-order explicit finite differences in the wall-normal direction). The computational domain has streamwise length of $L_x=4 \pi$ and spanwise length $L_z = 2 \pi$. The number of Fourier modes is $N_x=256$ in the streamwise direction and $N_z=256$ in the spanwise direction; the number of points in the wall-normal direction is $N_y=256$, unevenly spaced in order to decrease the grid size near the walls. The corresponding spatial resolution in the homogeneous directions is $\Delta x^+=6.5$ and $\Delta z^+=3.3$ (de-aliasing with the 3/2 rule is used); the wall-normal resolution increases from $\Delta y^+=0.5$ near the walls to $\Delta y^+=2.6$ at the centreline.

The equations of motion are advanced in time with a partially implicit approach, with an explicit three-substeps low-storage Runge--Kutta scheme combined with an implicit Crank--Nicolson scheme for the viscous terms. The size of the time step is $\Delta t = 0.02$, corresponding to an average value of CFL of 1.1.  

\subsection{Control strategies}

Two drag reducing flow control strategies are considered, namely the spanwise-oscillating wall or OW \citep{jung-mangiavacchi-akhavan-1992}, and the opposition control based on the wall-normal velocity component or VC \citep{choi-moin-kim-1994}. The OW control induces a spanwise wall movement resulting in a spanwise wall velocity distribution given by
\[
w_w(x,z,t)= W \sin \left( \frac{2 \pi}{T} t \right)
\]
where $W$ is the amplitude of the oscillation, and $T$ its period. 
VC control produces a distributed blowing and suction with the wall-normal velocity $v_w$ component at the wall opposing the same component in a wall-parallel plane at a prescribed wall distance $y = y_s$, according to 
\begin{equation*}
v_w(x,z,t)) = - v(x,y=y_s,z,t) .
\end{equation*}

Both control techniques are active, with the difference that OW requires a significant amount of energy to operate while the required control power for VC is marginal.
 
The control parameters are selected so that both techniques work around their optimum in the CPI sense, i.e. they achieve the maximum increase of bulk velocity $U_b$ with respect to the value $U_{b,0}$ in the reference uncontrolled channel. This corresponds for OW  to  $T=20.39$ and $W=0.1375$, or $T^+=125.5$ and $W^+=4.47$ in viscous units. For VC, the sensing plane is placed at $y_s=0.0653$ or $y_s^+=13.1$. Various quantities characterising the three simulations are summarised in table \ref{tab:cases}. To determine the values of $\gamma$, the control power per unit wetted area is computed for OW according to \cite{quadrio-ricco-2004} as $\aver{w_w \tau_z}$, where $\tau_z$ is the spanwise wall shear. $\power_c$ for the VC case is defined following \cite{choi-moin-kim-1994} as $\aver{p_w v_w + 0.5v_w^3}$, where $p_w$ is the pressure at the wall. The control is allowed to (locally or instantaneously) extract power from the flow, as the interest of the present work resides in the energy budget of the flow. Moreover, $\Pi_t$ is kept constant on a time-averaged sense, i.e. $\Pi_p$ is given by the total power minus the time-averaged value of $\Pi_c$. 

The calculations start from an initial condition where the flow has already reached statistical equilibrium for the specific controlled case, and are advanced for further $4000$ time units (corresponding to about 25000 viscous time units). During this time 200 flow fields are written to disk for the VC case; for the OW case, 200 flow fields are saved at 8 different phases of the oscillation, for a total of 1600 flow fields.

\begin{table}
\begin{tabular*}{1.0\textwidth}{@{\extracolsep{\fill}} l r r r r r r}
Case & $\Reyn$  & $Re_\tau$ & $Re_b$ & $\gamma$ & $U_b$  & $U_b/U_{b,0}$ \\

Ref  & 6500      & 199.7     &  3177  & 0        & 0.4887 & 1.000  \\ 
OW   & 6500      & 186.9     &  3267  & 0.09777  & 0.5026 & 1.028  \\ 
VC   & 6500      & 190.5     &  3474  & 0.00350  & 0.5345 & 1.094  \\ 
\end{tabular*}
\caption{Details of the CPI simulations. The table reports the adopted control strategy (see text for acronyms), the values of Reynolds numbers $\Reyn$, $Re_\tau$ and $Re_b$ (based on the power, friction and bulk velocity respectively), the fraction $\gamma$ of the control power, the bulk velocity $U_b$ and its ratio with the bulk velocity $U_{b,0}$ of the reference uncontrolled case. The results agree with the exact relation \eqref{eq:CPIconstraint} to within 0.2\%.}
\label{tab:cases}
\end{table}

It should be noted that the drag reduction rate $R$, i.e. the traditional figure of merit for flow control, is not a measure of improved energetic efficiency. $R$, defined as the percentage change of the skin friction coefficient $C_f=2\tau_w / \rho U_b^2$, is 17.2\% in the OW case, and 23.9\% in the VC case. These values are very similar (but not identical, since the comparison is made under a different condition) to the values of $R$ obtained at CFR at the condition of maximum net energy saving rate \citep{quadrio-ricco-2004,stroh-etal-2012}. Appendix A discusses the choices of figure of merit for evaluating flow control at CPI, and describes the link between $R$ and $U_b / U_{b,0}$, the figure of merit used in the present work.

%% file: boxes.tex
\section{Energy boxes}

\begin{figure}
\centering
\includegraphics[]{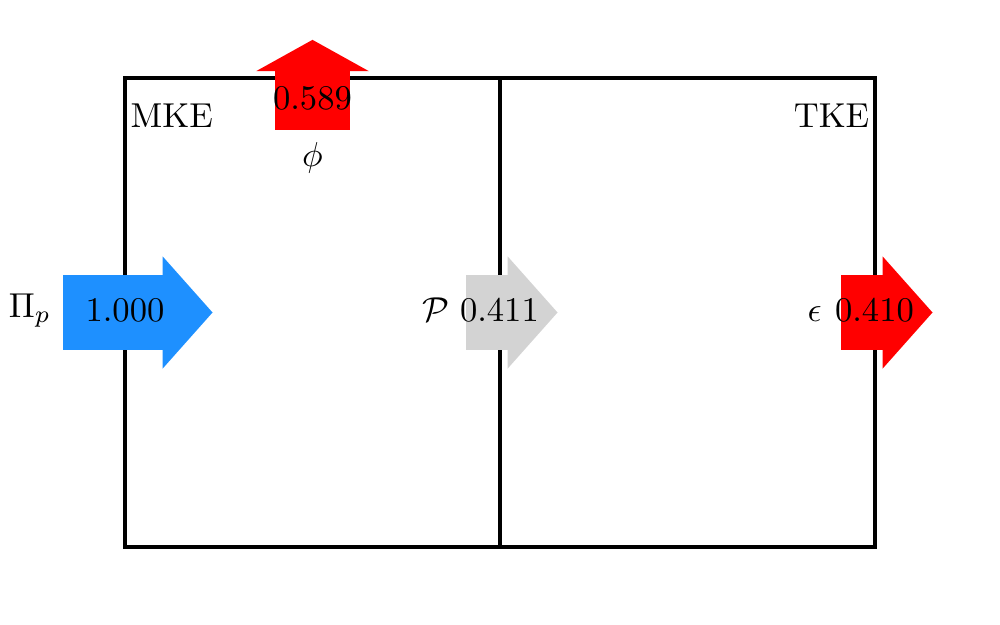}
\caption{Energy box in the reference case, with fluxes normalised by $\power_t = \power_p$. Owing to finite time average, the MKE budget has an unbalance of 0.05\%, and the TKE budget is unbalanced by 0.07\%. Here and in the following figures, the length of the arrows is proportional to the  flux magnitude.}
 \label{fig:ebox-ref}
\end{figure}
The energy box as introduced by \cite{ricco-etal-2012} is used to visualise the global balance of kinetic energy. The energy box for the uncontrolled reference case is shown in Fig.~\ref{fig:ebox-ref}, with numerical values for the energy fluxes scaled with the power input $\power_t = \power_p$. The MKE is either dissipated directly by the mean flow via the volume integral of the time-averaged viscous dissipation related to the mean velocity profile
\[
\dissU = \vaver{ \frac{1}{\Reyn} \left(\frac{\ud \aver{u}}{\ud y} \right)^2}
\]
or transferred to the TKE box via the production term
\[
\produ = \vaver{\uv \frac{\ud \aver{u}}{\ud y}}
\]
where 
\begin{equation}
\uv \equiv - \aver{u'v'} 
\label{eq:shear-stress}
\end{equation}
indicates the wall-normal distribution of the Reynolds shear stress. The TKE is either produced by $\produ$, and dissipated by the volume integral of the turbulent dissipation:
\[
\epsilon = \vaver{  \frac{1}{\Reyn} \frac{\partial u_i'}{\partial x_j} \frac{\partial u_i'}{\partial x_j} }
\]
where repeated indices imply summation.

Owing to statistical equilibrium, $\power_p = \dissU + \epsilon$ and $\produ = \epsilon$. The balance errors, due to the finite averaging time, are extremely small (less than $10^{-3}$) and demonstrate the quality of the dataset. It is interesting to mention that the simulations are carried out with a CPG condition for the spanwise component of the momentum equation, so that the power input to the spanwise part of the MKE is identically zero. However, given the finite averaging time, an extremely small residual spanwise mean velocity profile exists, which causes the related mean dissipation not to be identically zero, but still fully negligible (less than $10^{-6}$).

At the present value of Reynolds number, i.e. $\Reyn = 6500$, $59\%$ of $\Pi_t$ is dissipated by the mean field through $\dissU$, while the remaining $41\%$ is converted into TKE through $\produ$ and eventually dissipated through $\epsilon$. The relative importance of $\dissU$ and $\epsilon$ is a function of the Reynolds number, with the latter becoming dominant at higher $Re$ \citep{laadhari-2007,abe-antonia-2016}.

\begin{figure}
\centering
\includegraphics[]{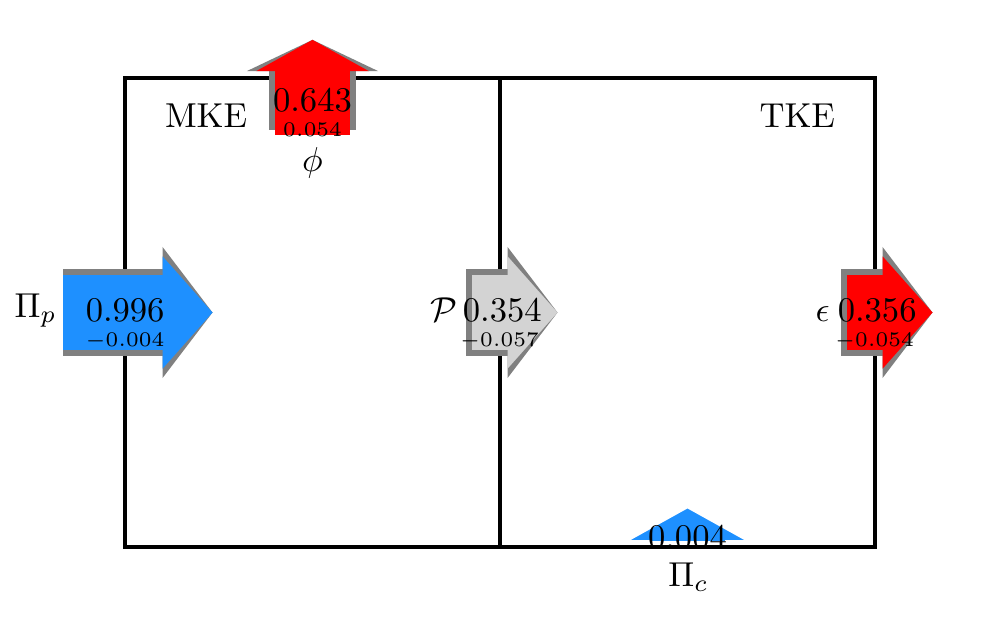}
\includegraphics[]{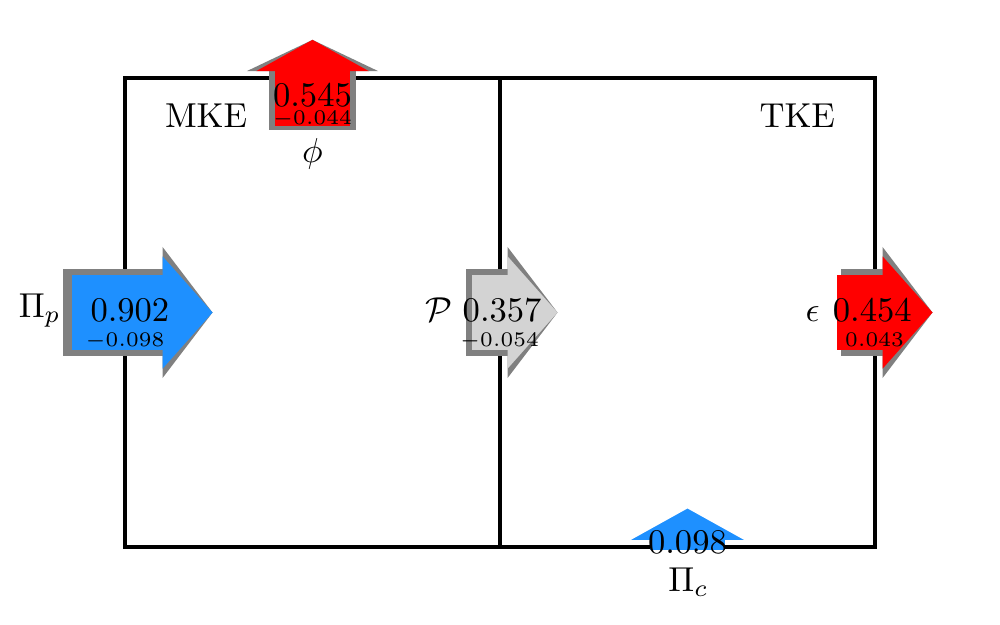}
\caption{Energy box in the VC (top) and OW (bottom) cases, with fluxes normalised by $\power_t$, and changes with respect to the reference case. For VC, the MKE budget has an unbalance of -0.10\%, and the TKE budget is unbalanced by 0.15\%. For OW, the MKE has an unbalance of -0.01\%, and the TKE is unbalanced by 0.2\%.}
\label{fig:ebox-CPI}
\end{figure}

Figure \ref{fig:ebox-CPI} shows the energy boxes for the VC and OW cases. As previously noted, in both cases $\power_c$ enters the system via the TKE box. While this is obvious for the zero-net mass flux VC, for OW within the classic Reynolds decomposition we do not distinguish the temporally-fluctuating quasi-laminar Stokes layer induced by wall oscillations \citep{quadrio-ricco-2004} from the chaotic turbulent fluctuations, but an alternative approach would be possible. In fact, $\power_c$ for OW can enter the MKE box for the time-varying but coherent spanwise velocity component, as done by \cite{ricco-etal-2012} and \cite{touber-leschziner-2012}, who adopted a triple decomposition for the velocity field. Our procedure can be easily modified to allow for the alternative approach without implications on the final findings.

The production term $\produ = \power_p - \dissU$ appears to consistently decrease when control is exerted, either because (VC case) $\dissU$ increases at nearly unchanged $\power_p$, or because (OW case) $\dissU$ decreases but $\power_p$ decreases even more, owing to the power requirements of the control action. The 
fraction of total power dissipated via $\dissU$ and $\epsilon$ exhibits opposite trends for the two types of control. For VC $\power_p$  only marginally decreases, because $\power_c$ is negligible; hence $\dissU$ increases by 0.054 compared to the uncontrolled flow, and $\epsilon$ decreases by the same amount. On the other hand, for the OW case, for which $\power_c$ is higher, the fraction of $\power_t$ dissipated by the mean flow decreases by 0.044.

These opposite trends are largely related to the different amount of $\power_p$ available for the two controls, and thus reflect the different control abilities to use $\power_c$ efficiently. In both cases the ratio $\dissU/\power_p$ increases, indicating that a larger part of the available pumping power is dissipated by the mean flow. However, $\dissU/\power_p$ is not a suitable indicator for energetically efficient flow control. One can imagine extreme cases in which $\dissU/\power_p$ increases because almost all available power is spent to run an highly-inefficient drag-reducing control which produces a small flow rate with the remaining small $\power_p$.

In order to quantitatively link the anticipated flow rate increase for a fixed $\power_t$ to the energy dissipation mechanisms and energy transfer rates, an enabling step involves giving prominence to the ultimate flow control target, i.e. reaching the laminar state. This will be achieved in the next Section, where an extension of the usual Reynolds decomposition is introduced.

%% file: decomposition.tex
\section{Extending the Reynolds decomposition}

\begin{figure}
\centering
\includegraphics[]{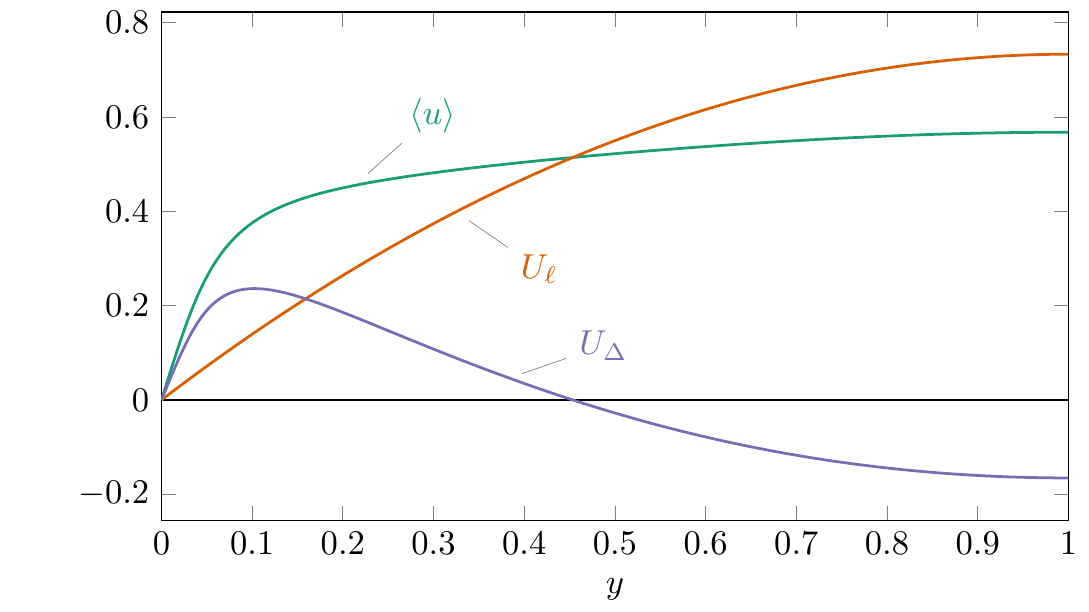}
\caption{Decomposition of the mean velocity profile $\aver{u}$ for the reference case into its laminar part $\Ulam$ and the resulting difference $\UDelta$.}
\label{fig:decompose-u}
\end{figure}

The most successful control would relaminarise the flow, thereby maximising the flow rate for the available pumping power. The control target, i.e. the achievement of a laminar flow, can be exposed in every quantity of interest by further decomposing the mean velocity in the usual Reynolds decomposition as:
\begin{equation*}
\aver{u}(y) = \Ulam(y) + \UDelta(y),
\end{equation*}
where $\Ulam$ is the laminar parabolic profile yielding the same $U_b$, and $\UDelta$ is the deviation of the actual mean velocity profile $\aver{u}$ from $\Ulam$. Fig.~\ref{fig:decompose-u} shows the profiles of $\aver{u}$, $\Ulam$ and $\UDelta$ for the reference case. By definition, the $\aver{u}$ and $\Ulam$ profiles possess the same bulk velocity, hence $\UDelta$ has zero average, i.e.
\begin{equation*}
\int_0^{1} \UDelta (y) \ud y = 0 .
\end{equation*}

Such decomposition is inspired by the work of \cite{eckhardt-grossmann-lohse-2007}, which discusses Taylor--Couette and Rayleigh--B\'enard flows. They introduce a ``convective'' (or ``wind'') dissipation rate, defined as the difference between the total dissipation rate and the dissipation rate of a laminar flow. It is worth noting that the present decomposition is as arbitrary as the classic Reynolds decomposition: the sum $\Ulam + \UDelta$ still amounts to the usual mean velocity profile. However, in contrast to $\aver{u}$, which is not a solution of the Navier--Stokes equation, $\Ulam$ is a possible state of the flow, occuring when a drag-reducing control achieves complete relaminarization. The fact that $\Ulam$ is analytically known is a key property that enables the following analysis, in which  kinetic energy transfer rates will be decomposed into separate components, which can in turn be related analytically to the Reynolds shear stress.  

\subsection{Three mean momentum equations}
As usual, the governing equation for the mean velocity profile $\aver{u}(y)$ is
\begin{equation}
\label{eq:U}
0 = G + \frac{\ud}{\ud y} \left( \frac{1}{\Reyn} \frac{\ud \aver{u}}{\ud y} + \uv \right) .
\end{equation}
where $r$ is the negative Reynolds shear stress \eqref{eq:shear-stress}, and $G$ corresponds to the negative mean streamwise pressure gradient.  
The governing equation for the laminar component $\Ulam$ is obtained by setting $\uv = 0$ in Eq.~(\ref{eq:U}):
\begin{equation}
\label{eq:U_lam}
0 = G_\ell + \frac{1}{\Reyn} \frac{\ud^2 \Ulam}{\ud y^2},
\end{equation}
where $-G_\ell$ is the streamwise pressure gradient required to achieve the same bulk mean velocity $U_b$ in a laminar flow.

One obtains the governing equation for $\UDelta$ by subtracting Eq.~\eqref{eq:U_lam} from Eq.~\eqref{eq:U}:
\begin{equation}
\label{eq:DU_1}
0 = G_\Delta + \frac{\ud}{\ud y} \left( \frac{1}{\Reyn} \frac{\ud \UDelta}{\ud y} + \uv \right),
\end{equation}
where $G_\Delta \equiv G - G_\ell$ is the increment of the pressure gradient due to the deviation of the mean velocity from a laminar parabolic one with the same $U_b$.

Obviously, the wall-normal distribution of Reynolds shear stress $\uv$ is essential to determine the mean velocity profile and the resulting control performance, as confirmed by Eqs.~\eqref{eq:U} and \eqref{eq:DU_1}.
In the following, mathematical expressions relating all budget terms in Fig.~\ref{fig:ebox} to $\uv$ and $\Reyn$ are derived. In order to do so, the integrals of the above momentum equations need first to be expressed in terms of $\uv$.

\subsection{Triple integration for $\aver{u}$: the FIK identity under CPI}

The original FIK identity was derived at CFR by \cite{fukagata-iwamoto-kasagi-2002}, and \cite{marusic-joseph-mahesh-2007} provided an analogous expression for CPG. In both cases, a weighted integral of the Reynolds shear stress, i.e. $\int (1-y) \uv \ud y $, describes either the additional drag generated by turbulence (at CFR) or the flow rate decrease due to turbulence (at CPG). For CPI, the influence of turbulence on the resulting flow rate can be captured based on the same weighted integral of $\uv$ as shown in the following.

First, Eq.~(\ref{eq:U}) is integrated in the $y$-direction from $0$ to $y$, yielding
\begin{equation*}
  (1-y)G = \frac{1}{\Reyn}\frac{\ud \aver{u}}{\ud y} + \uv.
  \label{eq:rofy}
\end{equation*}
One more integration in $y$ results in
\begin{equation*}
  \left( y-\frac{y^2}{2} \right) G = \frac{\aver{u}}{\Reyn} + \int_0^{y}\uv \ud y.
\end{equation*}
Finally, applying a third integration from $y = 0$ to $y = 1$ and integrating the rightmost term by parts provides the equation for the bulk mean velocity in the form
\begin{equation}
\label{eq:FIK_U_1}
U_b = \int_{0}^1 \aver{u} \ud y = \frac{G \Reyn}{3} - \alpha \Reyn .
\end{equation}
Here, $\alpha$ is defined as the weighted integral of $\uv$:
\begin{equation}
\label{eq:alpha}
\alpha = \int_0^{1} (1-y) \uv \ud y.
\end{equation}
Eq. ~(\ref{eq:FIK_U_1}) nicely shows that also for CPI the term containing $\alpha$ can be understood as the decrease of $U_b$ due to turbulence, since the laminar flow rate is given by $G \Reyn/3$. However, at CPI the streamwise pressure gradient $G$ changes with a change in $\alpha$. Therefore, in order to remove $G$ from Eq.~(\ref{eq:FIK_U_1}), we multiply it by $U_b$:
\begin{equation}
\label{eq:FIK_U_2}
U_b^2 = \frac{U_b G \Reyn}{3} - U_b \alpha \Reyn .
\end{equation}
and note that the quantity $U_b G$ is the pumping power input per unit area, which under the CPI condition is known by Eq.~\eqref{eq:power-p}. Substitution of this relation into Eq.~(\ref{eq:FIK_U_2}) leads to
\begin{equation}
\label{eq:FIK_U_3}
U_b^2  +  \alpha \Reyn U_b - (1 - \gamma) = 0 ,
\end{equation}
which can be solved to yield:
\begin{eqnarray}
\label{eq:Ub-alpha}
U_b &=& \frac{-\alpha \Reyn + \sqrt{ (\alpha \Reyn)^2 + 4(1 - \gamma)} }{2} \nonumber\\
&=& \frac{\alpha \Reyn}{2} \left\{ -1 + \sqrt{1 + \frac{4(1-\gamma)}{(\alpha \Reyn)^2}} \right\}
\end{eqnarray}
In a different form, this equation was already derived by \cite{hasegawa-quadrio-frohnapfel-2014} (see their Eq.(3.8) at p.99); it provides the relationship between $U_b$ and $\uv$, and is the FIK identity expressed for the CPI condition. Just like in the corresponding expressions for CFR and CPG, the Reynolds shear stresses appear in \eqref{eq:Ub-alpha} only via their weighted integral $\alpha$. 

\subsection{Triple integration for $\Ulam$}
Integrating Eq.~(\ref{eq:U_lam}) twice in the wall normal direction yields:
\begin{equation}
\label{eq:FIK_U_lam_1}
\Ulam(y) = \Reyn G_\ell \left( y-\frac{y^2}{2} \right).
\end{equation}
One more integration from $y = 0$ to $y = 1$ results in the expression for $U_b$:
\begin{equation}
\label{eq:Ub-Glam}
U_b = \int_{0}^{1} \Ulam(y) \ud y = \frac{\Reyn G_\ell}{3}.
\end{equation}

\subsection{Triple integration for $\UDelta$}
Integrating Eq.~(\ref{eq:DU_1}) twice in the wall-normal direction results in
\begin{equation}
\label{eq:FIK_DU_1}
\UDelta (y) = \Reyn G_\Delta \left( y - \frac{y^2}{2} \right) - \int_0^{y} \uv \ud y.
\end{equation}
By integrating once more from $y = 0$ to $y = 1$, one obtains
\begin{equation*}
\int_0^1 \UDelta (y) \ud y = 0 = \Reyn \left( \frac{G_\Delta}{3} - \int_0^1 (1-y) \uv \ud y \right).
\end{equation*}
Hence,
\begin{equation}
\label{eq:FIK_DU}
G_\Delta = 3 \int_0^1 (1-y) \uv \ud y = 3 \alpha.
\end{equation}

Equations \eqref{eq:Ub-Glam} and \eqref{eq:FIK_DU} are the contributions of the laminar and turbulent parts to the pressure gradient.

%% file: newboxes.tex
\section{Extending the energy box}

Having extended the Reynolds decomposition to account for the laminar component in the mean flow, we can now leverage it to redesign the energy box description of the channel flow. So far, expressions for the integrals of the corresponding velocity components containing the weighted integral $\alpha$ of the Reynolds shear stresses and the Reynolds number have been derived. In this section, every energy flux is related to the $\uv$ profile, eventually obtaining an extended version of the energy box.

The turbulent production term can be written as
\begin{equation*}
\produ = \int_0^1 \uv \frac{\ud \aver{u}}{\ud y} \ud y
=  \int_0^1 \uv \left( \frac{\ud \Ulam}{\ud y} + \frac{\ud \UDelta}{\ud y} \right) \ud y = \produ_\ell + \produ_\Delta,
\end{equation*}
where
\begin{equation}
\produ_\ell = \int_0^1 \uv \frac{\ud \Ulam}{\ud y} \ud y = 3 U_b \int_0^1 \left( 1 - y \right) \uv \ud y = 3 U_b \alpha
\label{eq:prodLam}
\end{equation}
is the turbulent production due to the laminar component and
\begin{equation*}
\produ_\Delta = \int_0^1 \uv \frac{\ud \UDelta}{\ud y} \ud y.
\end{equation*}
the one related to the deviation component. Note that $\produ_\ell$ is proportional to $\alpha$, the weighted integral of $\uv$ that also appears in the FIK identity. Following  the previously discussed interpretation of $\alpha$ in terms of ``losses'' associated with the presence of turbulence, $\produ_\ell$ can be understood as the fraction of $\power_p$ ``wasted'' by turbulence.
Similarly, $\produ_\Delta$  can be interpreted as the consequence of the existence of the Reynolds shear stresses, implying a deviation of the mean profile from the laminar profile.

\begin{figure}
\centering
\includegraphics[]{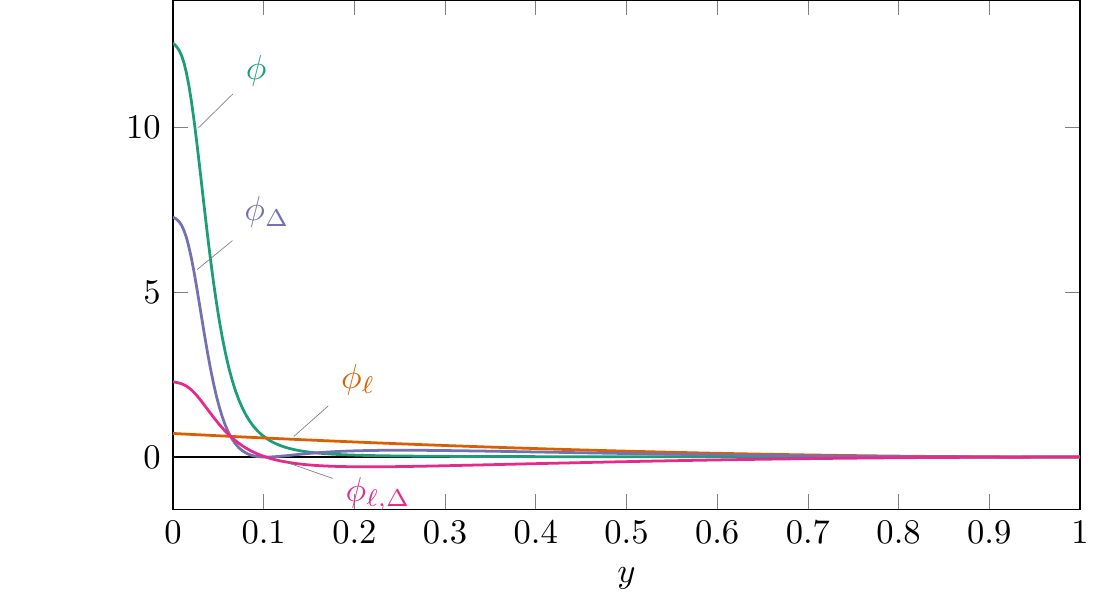}
\caption{Integrands of the dissipation terms $\dissU$, $\dissDelta$, $\dissLam$ and $\dissU_{\ell,\Delta}$ appearing in Eq.~\eqref{eq:decomposition-diss} in the reference case. The integral of the cross-term $\dissU_{\ell,\Delta}$ is zero.}
\label{fig:decompose-phi}
\end{figure}

Along the same lines, the mean dissipation can be decomposed as follows:
\begin{eqnarray}
\phi &=& \frac{1}{\Reyn}\int_0^{1} \left(\frac{\ud \aver{u}}{\ud y}\right)^2 \ud y
= \frac{1}{\Reyn} \int_0^{1} \left( \frac{\ud \Ulam}{\ud y} + \frac{\ud \UDelta}{\ud y} \right)^2 \ud y \nonumber\\
&=& \frac{1}{\Reyn}\int_0^{1} \left\{ \left(\frac{\ud \Ulam}{\ud y}\right)^2 + \left(\frac{\ud \UDelta}{\ud y} \right)^2
                                          + 2 \frac{\ud \Ulam}{\ud y} \frac{\ud \UDelta}{\ud y} \right\} \ud y = \dissLam + \dissDelta + \dissU_{\ell,\Delta}.
\label{eq:decomposition-diss}
\end{eqnarray}
The last term  $\dissU_{\ell,\Delta}$ vanishes since
\begin{equation*}
\int_0^1 \frac{\ud \Ulam}{\ud y} \frac{\ud \UDelta}{\ud y} \ud y
= \left. \UDelta \frac{\ud \Ulam}{\ud y} \right|_0^1 - \int_0^1 \UDelta \frac{\ud^2 \Ulam}{\ud y^2} \ud y = 0,
\end{equation*}
the last integral being zero because $\ud^2 \Ulam/\ud y^2 = - \Reyn G_\ell$ is a constant which can be taken out of the integral, and $\int_0^1 \UDelta \ud y = 0$. Therefore, the dissipation of the mean field is expressed as the sum
\begin{equation*}
\dissU = \dissLam + \dissDelta,
\end{equation*}
where 
\begin{equation*}
\dissLam = \frac{1}{\Reyn} \int_0^1 \left( \frac{\ud \Ulam }{\ud y} \right)^2 \ud y
\end{equation*}
is the energy dissipation rate associated with the laminar component, and 
\begin{equation*}
\dissDelta = \frac{1}{\Reyn} \int_0^1 \left( \frac{\ud \UDelta}{\ud y} \right)^2 \ud y
\end{equation*}
is the additional dissipation associated with the deviation component. 
The sum of $\dissDelta$ and $\epsilon$ is the share of $\power_t$ wastefully dissipated by turbulence, i.e. not used for creating a flow rate, for which only $\dissLam$ is necessary.
Fig.~\ref{fig:decompose-phi} shows the wall-normal distribution of the integrands of \eqref{eq:decomposition-diss}.
The advantage of decomposing $\dissU$ into $\dissLam$ and $\dissDelta$ is that $\dissLam$ possesses a unique relationship with $U_b$. Therefore, the only remaining issue is to understand how $\dissDelta$ changes in flows with increased energetic efficiency. 


We can also derive an interesting relationship between $\dissDelta$ and $\produ_{\Delta}$. Integrating Eq.~(\ref{eq:DU_1}) in the $y$ direction yields
\begin{equation}
\label{eq:DU_2}
G_\Delta (1-y) = \frac{1}{\Reyn} \frac{\ud \UDelta}{\ud y} + \uv .
\end{equation}
Multiplying by $\ud \UDelta / \ud y$ and integrating in $y$, the following equation is obtained:
\begin{equation*}
\int_0^1 G_\Delta (1-y) \frac{\ud \UDelta}{\ud y} \ud y
= \frac{1}{\Reyn}\int_0^1 \left(\frac{\ud \UDelta}{\ud y}\right)^2 \ud y
 + \int_0^1 \uv \frac{\ud \UDelta}{\ud y} \ud y
= \dissDelta + \produ_\Delta.
\end{equation*}
The left-hand-side is zero, because
\begin{equation*}
G_\Delta \int_0^1(1-y) \frac{\ud \UDelta}{\ud y} \ud y = \left. (1-y) \UDelta \right|_0^1 + \int_0^1 \UDelta \ud y = 0.
\end{equation*}
Hence, the following identity is obtained
\begin{equation}
\label{eq:PD-phiD}
\produ_\Delta = - \dissDelta \leq 0,
\end{equation}
which emphasizes that the production $\produ_\Delta$ is always a source for MKE and a sink for TKE. 
This is reasonable, since the laminar profile is the one with minimum energy dissipation for a given flow rate. Therefore, any deviation from the laminar profile corresponds to additional energy losses, expressed by $\dissDelta$.  

Moreover, Eq.~(\ref{eq:PD-phiD}) indicates that both $\produ_\Delta$ and $\dissDelta$ cancel each other in the MKE budget, which becomes:
\begin{equation}
  \power_p =  \dissLam + \produ_\ell. 
\label{eq:def_PP}
\end{equation}


\begin{figure}
\centering
\includegraphics[]{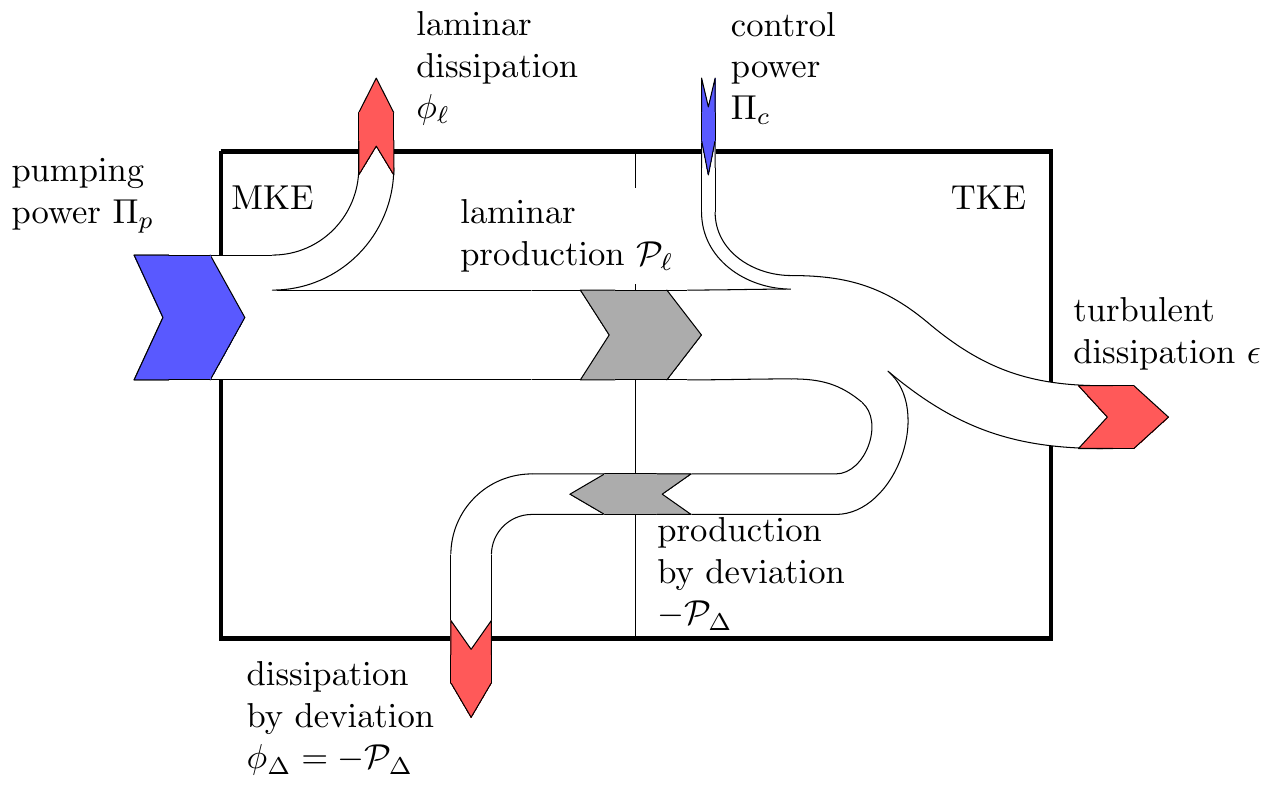} 
\caption{Updated energy box when the mean velocity profile $\aver{u}$ is decomposed into laminar $\Ulam$ and the deviation $\UDelta$ from the laminar profile.}
\label{fig:ebox-extended-sketch}
\end{figure}

In summary, the original energy box sketched in Fig.~\ref{fig:ebox} can now be extended as shown in Fig.~\ref{fig:ebox-extended-sketch}.
The dissipation of the mean velocity and the turbulent production are decomposed into their contributions from the laminar profile $\Ulam$ and the deviation $\UDelta$ from the laminar profile. The turbulent production $\produ_\Delta$ due to $\UDelta$ is always negative, so that the arrow is pointing in the opposite direction, i.e. from TKE to MKE. 

\subsection{Energy fluxes as a function of $\uv$}
Every energy flux appearing in Fig.~\ref{fig:ebox-extended-sketch} is expressed as a function of the Reynolds shear stresses and of the Reynolds number in the following.

\subsubsection{Input powers $\power_p$ and $\power_c$}
By definition, the pumping and control powers are expressed by:
\begin{equation}
\power_p = \frac{3(1 - \gamma)}{\Reyn}; \qquad \power_c = \frac{3 \gamma}{\Reyn} .
\label{eq:power-pc}
\end{equation}

\subsubsection{Laminar dissipation $\dissLam$}
Integration of Eq.~(\ref{eq:U_lam}) in $y$ leads to
\begin{equation*}
G_\ell (1-y)  =  \frac{1}{\Reyn} \frac{\ud \Ulam}{\ud y}.
\end{equation*}
Multiplying by $\ud \Ulam/\ud y$ and integrating from $y = 0$ to $y = 1$, results in
\begin{equation*}
\int_0^1 G_\ell (1-y)\frac{\ud \Ulam}{\ud y} \ud y  =  \int_0^1 \frac{1}{\Reyn} \left(\frac{\ud \Ulam}{\ud y}\right)^2 \ud y = \dissLam .
\end{equation*}
Substituting $\Ulam$ with Eq.~(\ref{eq:FIK_U_lam_1}) on the left-hand-side gives
\begin{equation}
\dissLam =  \Reyn G^2_\ell \int_0^1 (1-y)^2 \ud y = \frac{\Reyn G^2_\ell}{3} = \frac{3U_b^2}{\Reyn},
\label{eq:phi_lam-noFIK}
\end{equation}
where Eq.~(\ref{eq:Ub-Glam}) has been used for the final expression.
Using Eqs.~\eqref{eq:FIK_U_3} and \eqref{eq:Ub-alpha} yields:
\begin{eqnarray}
\label{eq:phi_lam}
\dissLam &=& \frac{3}{\Reyn} \left\{ - \alpha \Reyn U_b + (1 - \gamma)\right\} \nonumber\\
&=& \frac{3}{\Reyn} \left\{ \frac{(\alpha \Reyn)^2}{2} \left( 1 - \sqrt{1 + \frac{4(1-\gamma)}{(\alpha \Reyn)^2}} \right)
                                      + (1 - \gamma) \right\} .
\end{eqnarray}
Note that $\uv$ enters this expression only through its weighted integral $\alpha$. 

\subsubsection{Laminar production $\produ_\ell$}
\label{sec:Plam}
The turbulence production $\produ_\ell$ related to the laminar component (which is indeed an oxymoron) is given by
\begin{eqnarray}
\produ_\ell &=& \int_0^1 \uv \frac{\ud \Ulam}{\ud y} \ud y
= \Reyn G_\ell \int_0^1 \uv (1-y) \ud y \nonumber\\
&=& 3\alpha U_b
= \frac{3}{\Reyn} \frac{(\alpha \Reyn)^2}{2} \left\{ -1 + \sqrt{1 + \frac{4(1-\gamma)}{(\alpha \Reyn)^2}} \right\} .
\label{eq:Plam}
\end{eqnarray}
Again, $\produ_\ell$ depends on $\uv$ via $\alpha$ only. We note that $\produ_\ell$ can assume negative values in the rare cases when $\alpha<0$ which corresponds to an inverse sign of the Reynolds shear stress. This is e.g. achieved by streamwise travelling waves of blowing and suction \citep{min-etal-2006}. In this control case sublaminar drag is reported at CFR which would correspond to $\dissLam/\power_p>1$ at CPI, see Eq.~\eqref{eq:def_PP}. It should be kept in mind that, by definition, it is impossible to obtain $\dissLam/\power_t>1$; i.e. the controlled flow cannot produce a flow rate higher than the one achieved when all available power is spent for pumping a laminar flow.

\subsubsection{Production $\produ_\Delta$ and dissipation $\dissDelta$}
Using Eq.~(\ref{eq:DU_2}) one obtains
\begin{eqnarray}
\dissDelta &=& \int_0^1 \frac{1}{\Reyn} \left(\frac{\ud \UDelta}{\ud y}\right)^2 \ud y
= \Reyn \int_0^1 \left\{ G_\Delta (1-y) - \uv \right\}^2 \ud y \nonumber\\
&=& \Reyn \int_0^1 \left\{ G_\Delta^2 \left(1-y\right)^2 - 2 G_\Delta (1-y)\uv + \uv^2 \right\} \ud y \nonumber\\
&=& \Reyn \left\{ \frac{G_\Delta^2}{3} - 2 G_\Delta \alpha + \int_0^1 \uv^2 \ud y \right\} \nonumber\\
&=& \Reyn \left( \beta - 3 \alpha^2 \right),
\label{eq:phi_delta}
\end{eqnarray}
where Eq.~(\ref{eq:FIK_DU}) is used and $\beta$ is defined as
\begin{equation}
\beta = \int_0^1 \uv^2 \ud y.
\label{eq:beta}
\end{equation}
Hence, $\dissDelta$ is again expressed as a function of $\uv$, but in addition to the weighted integral $\alpha$ the term $\beta$ also appears, corresponding to the integral of $\uv^2$.

According to \eqref{eq:PD-phiD}, the turbulent production $\produ_\Delta$ is obtained as
\begin{equation}
\produ_\Delta = - \dissDelta = \Reyn \left( 3 \alpha^2 - \beta \right) \leq 0.
\label{eq:P_Delta}
\end{equation}
Therefore, $\beta \geq 3 \alpha^2$. While $\alpha$ can switch sign, $\beta$ is always positive which is consistent with the fact that any deviation from the laminar profile corresponds to energetic losses.

\subsubsection{Turbulent dissipation $\epsilon$}
The turbulent dissipation is obtained from the energy balance of TKE shown in the right box in Fig.~\ref{fig:ebox-extended-sketch}:
\begin{eqnarray}
\epsilon &=& \produ_\ell + \produ_\Delta + \power_c \nonumber\\
&=& \frac{3}{\Reyn}
  \left\{  \frac{(\alpha \Reyn)^2}{2} \left( 1 + \sqrt{1 + \frac{4(1-\gamma)}{(\alpha \Reyn)^2}} \right)
          - \frac{\beta \Reyn^2}{3}+ \gamma \right\} .
  \label{eq:epsilon}
\end{eqnarray}
Here, Eqs.~\eqref{eq:power-pc}, \eqref{eq:Plam} and \eqref{eq:P_Delta} are used.

\subsection{The extended energy box}

\begin{figure}
\centering
\includegraphics[]{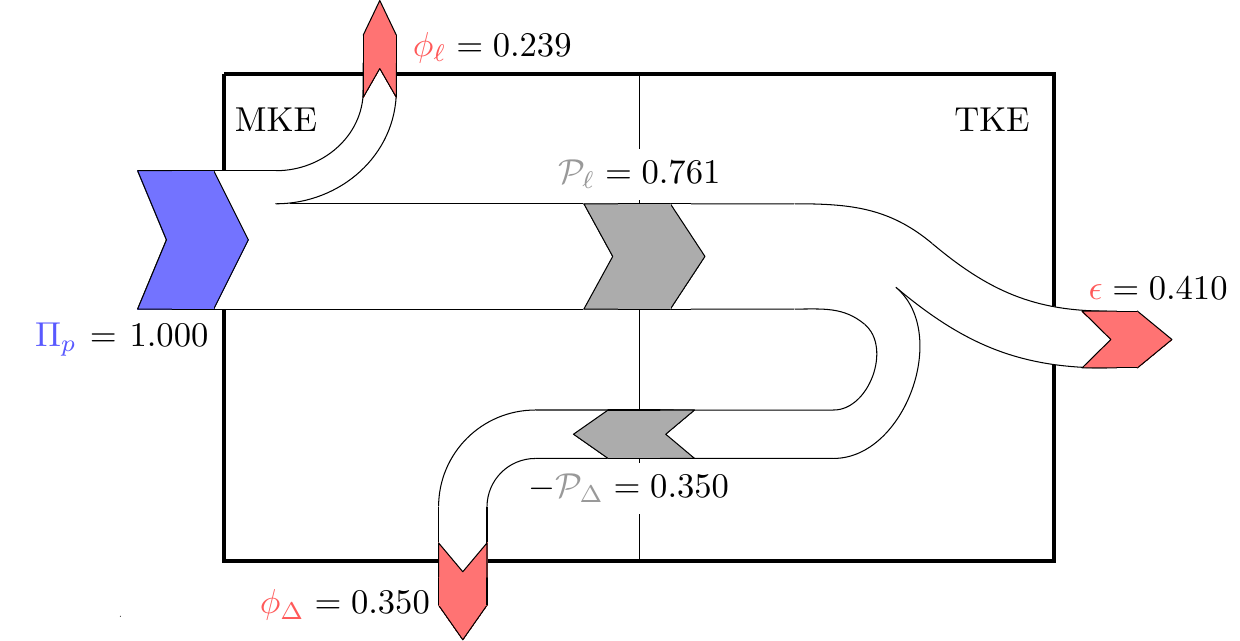}
\caption{Extended energy box for the reference case, with fluxes normalised with $\power_t=\power_p$.}
\label{fig:ebox-extended-nocontrol}
\end{figure}

Figure \ref{fig:ebox-extended-nocontrol} shows the extended energy box for the reference flow without drag reduction. At this low value of the Reynolds number, at which $\dissU$ is known \citep{laadhari-2007} to overwhelm $\epsilon$, this decomposition highlights that $\dissLam$ is about one fourth of the total power, and is comparable to $\dissDelta$. The share of $\power_t$ that is not being used to produce a flow rate (it should be remembered that the only velocity profile contributing to the flow rate is $\Ulam$) is $\dissDelta + \epsilon = \produ_\ell$. This power depends on $\uv$ only via $\alpha$, and can be considered as power wasted to produce turbulence; here it is approximately 76\% of the total power, a fraction that is expected to increase with $Re$.

How these fluxes vary with $Re$ can be examined by resorting to empirical formulas expressing how $\dissU^+$ and $\epsilon^+$ change with $Re$; such formulas are for example discussed by \cite{abe-antonia-2016}.  Appendix B reports this analysis, which leads to a new and possibly improved relationship between $Re_b$ and $Re_\tau$.

\begin{figure}
\centering
\includegraphics[]{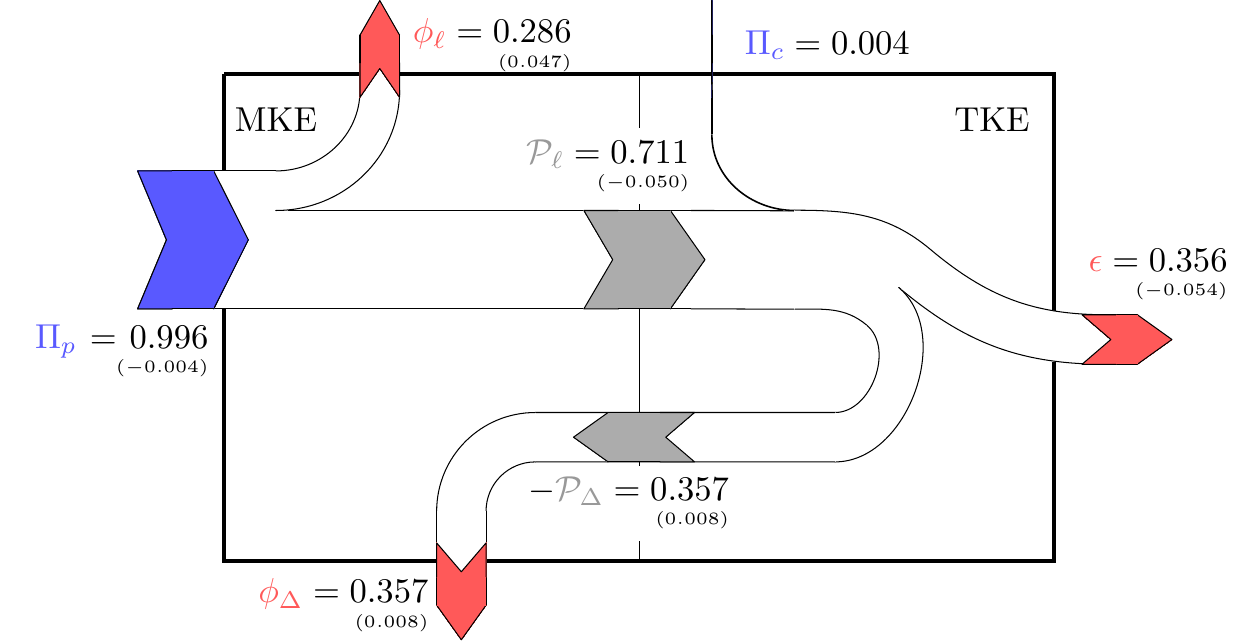}\\[25pt]
\includegraphics[]{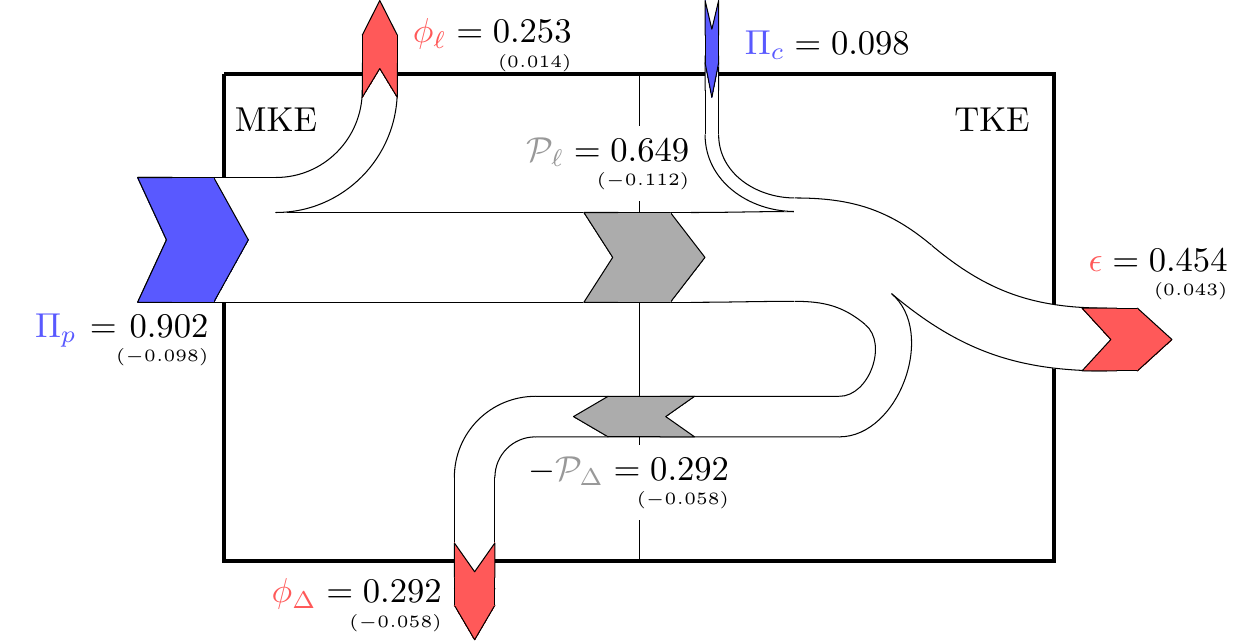}
\caption{Extended energy box in the VC (top) and OW (bottom) cases, with fluxes normalised by $\power_t$. Changes from the reference case are shown in parentheses.}
\label{fig:ebox-extended-control}
\end{figure}
\begin{table}
\begin{tabular*}{1.0\textwidth}{@{\extracolsep{\fill}} l r r r r r r r}
Case & $\power_p$ & $\dissLam$ & $\dissDelta$ & $\produ_\ell$ & $-\produ_\Delta$  & $\power_c$ & $\epsilon$ \\

Ref  & 1.000      & 0.239      &  0.350       & 0.761         & 0.350             & -          & 0.410      \\ 
VC   & 0.996      & 0.286      &  0.357       & 0.711         & 0.357             & 0.004      & 0.356      \\ 
OW   & 0.902      & 0.253      &  0.292       & 0.694         & 0.292             & 0.098      & 0.454      \\

\end{tabular*}
\caption{Summary of the fluxes in the extended energy box for the reference, VC and OW case. All fluxes are normalised by $\power_t$. \label{tab:sankey}}
\end{table}

Figure \ref{fig:ebox-extended-control} shows the extended energy box for the considered control cases. The numerical values of all fluxes are summarised in Table \ref{tab:sankey}.

Owing to the CPI approach, $\power_p$ is decreased when a positive $\power_c$ is present for active control. Being proportional to $U_b^2$, $\dissLam$ increases when the control is successful, leading to an increased flow rate. However, the increase of $\dissLam$ is much larger for VC than for OW, for which the ability to increase $U_b$ is hindered by the reduced pumping power available under CPI.

The term $\produ_\ell$ is observed to decrease in both the controlled cases. In fact, the change of $\produ_\ell$ is prescribed by the power budget of the left panel of the extended energy box, which reads
\[
\produ_\ell = \power_p - \dissLam
\]
since $\dissDelta = - \produ_\Delta$. Equivalently, this budget can be rewritten by using \eqref{eq:Plam}, \eqref{eq:power-p} and \eqref{eq:phi_lam-noFIK} as:
\[
3 \alpha U_b = \frac{3(1-\gamma)}{\Reyn} - \frac{3U_b^2}{\Reyn}.
\]
In the present context $\power_p$ must decrease when active control is on, and $\dissLam$ must increase when control is successful, so that their difference $\produ_\ell$ must decrease. This result has the interesting implication that the product $\alpha U_b$ must decrease as well, which is not obvious since $U_b$ increases while $\alpha$ decreases. This implication is related to the flow rate increase at CPI being bounded by $U_b \leq 1$, whereas $\alpha$ can theoretically even drop to negative values, as mentioned in \S\ref{sec:Plam}.

The sign change of $\produ_\Delta$ (and $\dissDelta$) is however undefined. In fact, in our two examples $\produ_\Delta$ slightly increases for VC but decreases for OW. This is a consequence of the presence, see e.g. Eq.~\eqref{eq:P_Delta}, of both $\alpha$ and $\beta$ in their definition, but with opposite signs, and highlights differences in $r(y)$ between the two controlled cases. This observation reveals that a reduction of $\produ_\Delta$ is not sufficient for achieving energetically successful flow control, despite $\produ_\Delta$ being always detrimental to achieving higher flow rate. In the present case, for instance, OW reduces $\produ_\Delta$, while VC fails in doing so. 

As a consequence, also the sign of the last flux $\epsilon$ is in general undetermined, and in fact in our two examples $\epsilon$ decreases for VC and increases for OW. However, in the specific OW case, we know \citep{quadrio-ricco-2011} that the wall oscillations generate a spanwise Stokes flow that, even in a turbulent channel flow, closely resembles the laminar Stokes solution. Hence, nearly the whole $\power_c$ (the precise figure is 97\% in the present case) is dissipated directly by the spanwise Stokes layer, instead of the small-scale fluctuating turbulent field. If the contribution of the Stokes layer is removed, $\epsilon$ decreases in the OW case, too. 

\subsection{Relationship between flow rate increase, $\dissU$ and $\epsilon$}
All the terms featuring in the extended energy box have been related to the profile $\uv$ of the Reynolds shear stresses. We can then return to our original goal, and discuss how the dissipation rates of the mean and fluctuating fields are affected by flow control techniques intended to increase flow rate. Thanks to the CPI constraint, the sum $\dissU +\epsilon$ is always unchanged, so that considering one of the two terms is sufficient. From the simple energy box sketched in Fig.~\ref{fig:ebox}, the global energy balance indicates that:
\[
\dissU = \frac{3}{\Reyn} - \epsilon = \power_t - \epsilon .
\]
Using Eq.~\eqref{eq:phi_lam} and \eqref{eq:phi_delta}, The dissipation $\dissU$ is expressed by
\begin{eqnarray}
\dissU &=& \dissLam + \dissDelta =
\frac{3}{\Reyn} \left\{ \frac{(\alpha \Reyn)^2}{2} \left( 1 - \sqrt{1 + \frac{4(1-\gamma)}{(\alpha \Reyn)^2}} \right)
                                      + (1 - \gamma)\right\}
+ \Reyn \left( - 3 \alpha^2 + \beta \right)        \nonumber\\
&=&
\frac{3}{\Reyn} \left\{ \frac{(\alpha \Reyn)^2}{2} \left( -1 - \sqrt{1 + \frac{4(1-\gamma)}{(\alpha \Reyn)^2}} \right)
                                      + \frac{\beta \Reyn^2}{3}+ (1 - \gamma)\right\} .
\label{eq:phi_fin}
\end{eqnarray}
which only contains $\alpha$, $\beta$, $\gamma$ and $\Reyn$.

Although the flow rate increase is uniquely determined by $\alpha$ as shown by Eq.~(\ref{eq:Ub-alpha}), both $\dissU$ and $\epsilon$ additionally involve $\beta$. Its definition \eqref{eq:alpha} shows that $\alpha$ is the correlation between $\uv$ and $(1-y)$, whereas Eq.~\eqref{eq:beta} shows that $\beta$ is the L2 norm of $\uv$. As long as the distribution $\uv$ is unspecified, $\alpha$ and $\beta$ are unknown. Therefore, the relationship between flow rate increase and the changes in $\dissU$ and $\epsilon$ is also unknown --- knowing it would be equivalent to solving the closure problem of turbulence.

\begin{figure}
\centering
\includegraphics{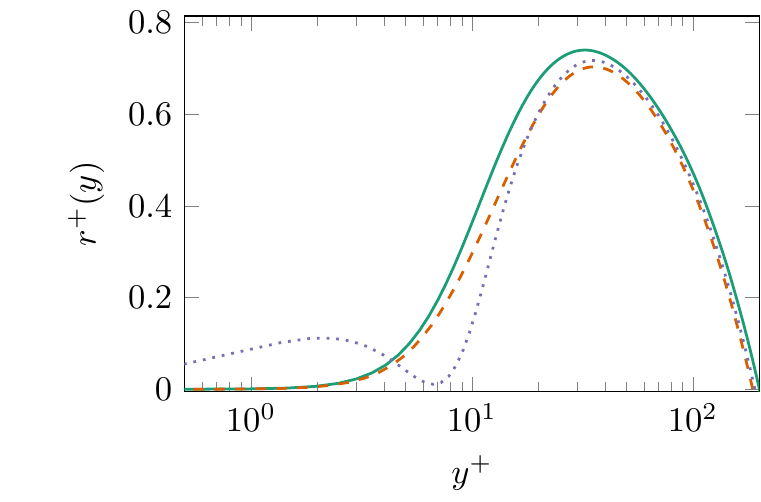}
\includegraphics{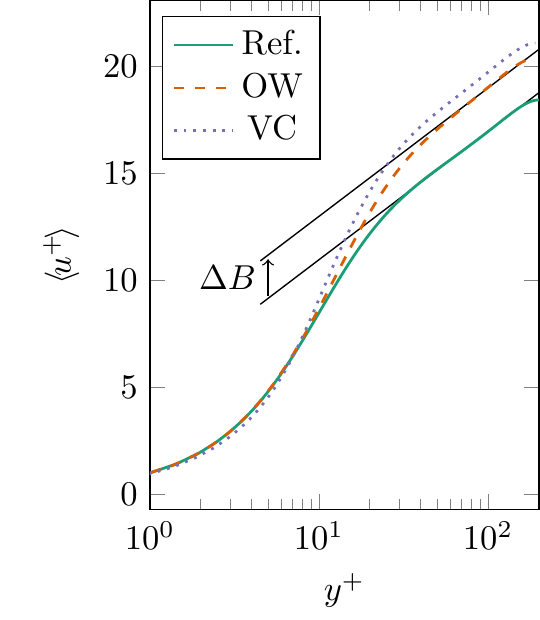}
\caption{Left: changes in $\uv$ induced by OW and VC control. Right: mean velocity profiles for the reference case and the controlled OW and VC cases. All quantities are plotted in actual wall units.}
\label{fig:profiles}
\end{figure}

One can however consider the space of possible controls, and assess how $\dissU$ and $\epsilon$ change in controlled flows, by exploring a range of values for $\gamma$ and parametrized changes of the profile $r_0(y)$ of the Reynolds shear stresses of the uncontrolled flow. Parametrising $r_0(y)$, however, is quite an arbitrary step. Fig.~\ref{fig:profiles} (left) shows how $r_0(y)$ changes from the reference flow in the VC and OW controlled cases. The changes are different in nature: the OW profile is mostly rescaled while the VC profile behaves differently near the wall, with a secondary zero point in the flow \citep[the position of the so-called virtual wall,][]{choi-moin-kim-1994}. However, in both cases the control-induced modifications of $r(y)$ yield very similar effects upon $\aver{u}(y)$, as illustrated in Fig.~\ref{fig:profiles} on the right. The controlled flows exhibit an upward shift $\Delta B^+$ (in actual viscous units) of the logarithmic portion of the mean velocity profile, defined as 
\begin{equation}
\aver{u^+} = \frac{1}{\kappa}\log y^+ + B + \Delta B   \,,
\label{eq:loglaw}
\end{equation}
in which $\kappa$ is the von K\'arm\'an constant and $B$ an additive constant. For the OW and VC cases the measured value of $\Delta B^+$ is 2.03 and 2.76 respectively. 

\begin{figure}
\centering
\includegraphics{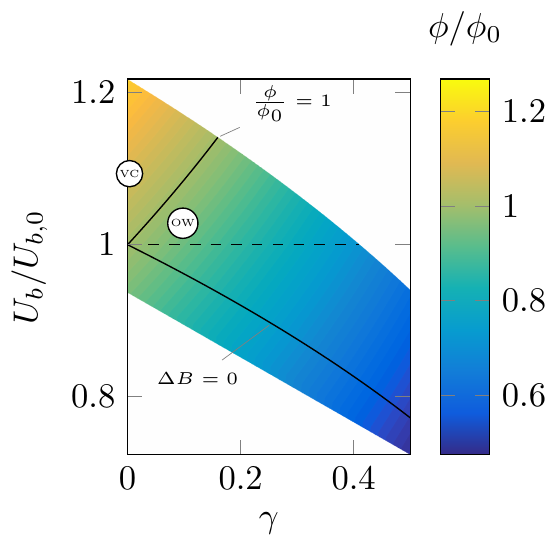}
\includegraphics{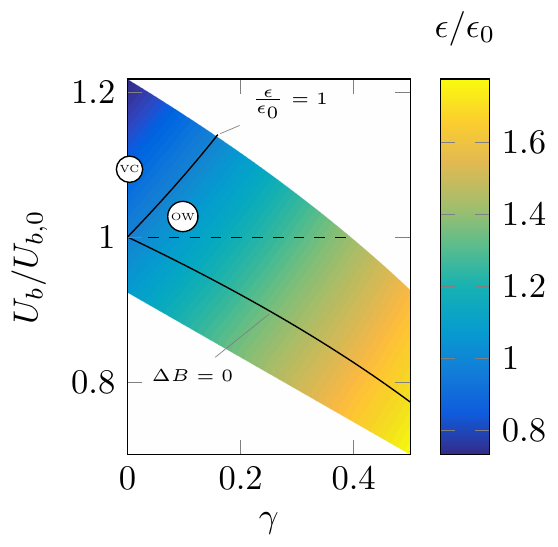}
\caption{Changes in $\dissU/\dissU_0$ (left) and $\epsilon/\epsilon_0$ (right),for the uncontrolled flow at $Re_\tau=200$, as a function of $\gamma$ and the flow rate increase $U_b/U_{b,0}$ occurring because of a shift $\Delta B$ in the logarithmic portion of the mean velocity profile at CPI. The symbols indicate the VC and OW cases.}
\label{fig:changes}
\end{figure}

We thus devise a procedure that uses the approximate analytical form of the mean velocity profile proposed by \cite{spalding-1961} as a baseline over which the parameter $\Delta B^+$ is varied within the range $\Delta B^+ = -2$ to $\Delta B^+ = 7$ to mimic the effect of wall-based flow control. The log layer constant are set to the values of $B=4.48$ and $\kappa=0.392$ as proposed by \cite{luchini-2018}. (The same work reviews the existing approximations for $\aver{u}(y)$ and proposes an elegant semi-empiric formula, which restores universality of the log layer constants among different canonical internal flows. Unfortunately, this expression could not be adopted here as it does not easily accommodate a variable $\Delta B^+$. However, neither the specific expression of $\aver{u}(y)$ nor the way changes of the log layer constants are implemented affect the following discussion.) $Re_\Pi = 6500$ is imposed, which corresponds to $Re_\tau=200$ for the reference channel ($\Delta B^+=0$). At CPI for this value of $Re_\Pi$, the range of $\Delta B^+$ corresponds to values of $U_b / U_{b,0}$ between 0.724 to 1.218.

For each $\Delta B^+$, the value of $Re_\tau$ is adjusted such that condition \eqref{eq:CPIconstraint} is satisfied. The distribution of $r(y)$ is then computed according to Eq.~\eqref{eq:U}, from which $\alpha$ and $\beta$ are evaluated and then eventually inserted into formula \eqref{eq:phi_fin}. This is certainly a simplified and approximated procedure, which is not expected to hold for any type of control technique. However, it is general enough to be valid for the two control strategies considered in the present work, which are representative of wall-based control.

Figure \ref{fig:changes} depicts the changes in $\dissU$ and $\epsilon$ predicted via this procedure. The vertical axis shows the flow rate $U_b/U_{b,0}$ that occurs in presence of a shift $\Delta B$ at CPI, and the horizontal axis is the control effort $\gamma$. The DNS results of the two flow control techniques discussed before, namely OW and VC, are also included in the plot. The region where $U_b/U_{b,0}>1$ corresponds to successful control under CPI. The model nicely visualizes and generalizes (within the simplifications indicated above) the result obtained for OW and VC: when active control techniques with variable $\gamma$ are considered at CPI, the way energy is dissipated in an energetically more efficient system does not imply a unique trend in $\dissU$ (or $\epsilon$). However, along the vertical axis ($\gamma=0$, passive control) an increase of $\dissU$ (or a decrease of $\epsilon$) is always related to a successful control, although, strictly speaking, this is only true for the present model. 

In the lower half of the plots in figure \ref{fig:changes} the flow rate achieved at constant $\Pi_t$ decreases. The line $\Delta B = 0$ marks flow states where flow control does not modify $\uv$ and hence the mean velocity profile. Here the control effect is only a reduction of $\Pi_p$. In the region above the line $\Delta B = 0$ but below $U_b/U_{b,0}=1$, the control increases the flow rate compared to a canonical channel driven at $\Pi_p=(1-\gamma)\Pi_t$, i.e. it interacts positively with turbulence and reduces $\alpha$. However, owing to its power cost, it does not go above $U_{b,0}$, corresponding to flow control techniques that at CFR produce drag reduction but no net energy savings. Lastly, below the line $\Delta B = 0$ the control effect on turbulence is detrimental. Such a scenario is observed for example for the travelling waves of spanwise wall velocity \citep{quadrio-ricco-viotti-2009} in their drag-increasing regime, and at $\gamma=0$ is typical of most rough surfaces.

\begin{figure}
\centering
\includegraphics{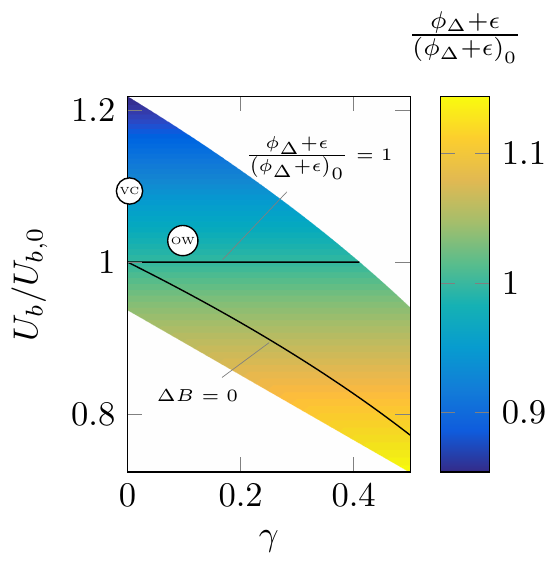}
\includegraphics{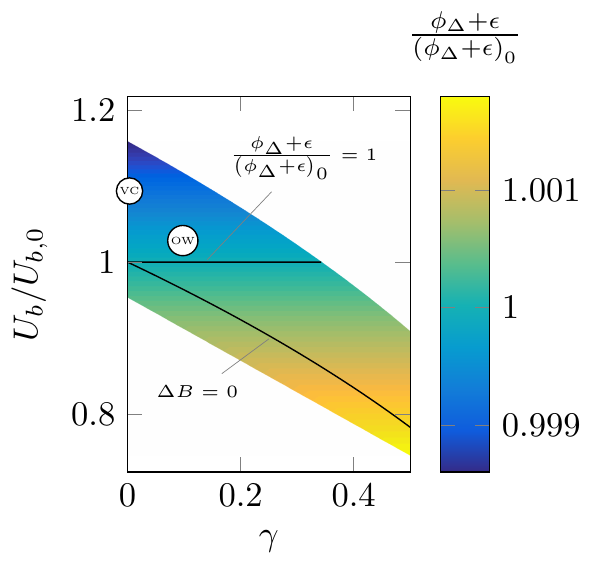}
\caption{Left: changes in $(\dissDelta + \epsilon)/(\dissDelta + \epsilon)_0$ (left) for the uncontrolled flow at $Re_\tau=200$, as a function of $\gamma$ and the flow rate increase $U_b/U_{b,0}$ occurring because of a shift $\Delta B$ in the logarithmic portion of the mean velocity profile at CPI. The symbols indicate the VC and OW cases. Right: changes in $\dissU/\dissU_0$ at $Re_\tau=20000$.}
\label{fig:changes2}
\end{figure}

Figure \ref{fig:changes2} (left) shows the distribution of  $\dissDelta  + \epsilon$, i.e. the fraction of total power wastefully dissipated by turbulence, as a function of $U_b/U_{b,0}$ and $\gamma$. It can be seen that in contrast to  $\dissU$ or $\epsilon$
this quantity does not depend on $\gamma$, but a decrease of $\dissDelta  + \epsilon$ is directly linked to an increase of $U_b/U_{b,0}$, the indicator of successful control. 

The simple model presented above can easily be extended to higher Reynolds numbers in order to verify the robustness of the present evidence against changes of $Re$. In Fig.\ref{fig:changes2} the same range of $\Delta B^+$ is considered at the increased Reynolds number of $Re_\Pi\approx 853000$, which yields  $Re_\tau=20000$ for a reference channel. This value of $Re$ is large enough for the various budget terms to reach their asymptotic behaviour (see Appendix B). Since the link between $\Delta B$ and the induced change in flow rate is $Re$-dependent, the range  $ -2 \leq \Delta B^+ \leq 7$ corresponds to smaller variations of flow rate at this larger $Re$, with values of $U_b / U_{b,0}$ ranging between 0.745 and 1.158. This reflects the known decrease of control efficiency with increasing Reynolds number for "similar" wall-based controls \citep{gatti-quadrio-2016}. The corresponding model results reveal that, also at high Reynolds number, changes in  $\dissU$ or $\epsilon$ can be of either sign. Figure \ref{fig:changes2} (right) shows the distribution of  $\dissDelta  + \epsilon$ at the high Reynolds number, that remains independent from $\gamma$, thus confirming that $\dissDelta  + \epsilon$ has a consistent trend for energetically efficient flow control at CPI, whereas $\epsilon$ alone can either increase or decrease.  When comparing the left and right part of Fig.\ref{fig:changes2} it is also interesting to note that the relative change in $\dissDelta  + \epsilon$ for the controlled flows is two orders of magnitude smaller, indicating that at high Reynolds number basically all pumping power is wasted by turbulence (see Appendix B).

%% file: conclusion.tex
\section{Concluding remarks}

The present work has considered how flow control aimed at improving the  energetic efficiency of turbulent channel flows changes the energy fluxes, especially the energy dissipation rates $\dissU$ and $\epsilon$ of the mean and fluctuating velocity fields. 
Such an analysis is difficult when based on data obtained at CFR or CPG, where flow control brings about an inherent change of the total power input: the total viscous dissipation rate $\dissU + \epsilon$ of kinetic energy changes whenever control is applied. Therefore, the Constant Power Input (CPI) concept has been selected to compare energy transfer rates between controlled and uncontrolled flow. With CPI the total viscous dissipation rate is fixed, and the success of control implies an increase of the flow rate for the same value of $\dissU + \epsilon$, which is equivalent to the total power entering the flow system. In order to keep the total power constant, the power $\power_c$ consumed by the considered active flow control techniques is subtracted from the available pumping power $\power_p$. 

The obtained results show that  $\dissU$ or $\epsilon$ undergo changes of either sign in a successfully controlled flow, depending on $Re$ and the particular way in which the control strategies modify $r(y)$. Therefore, these quantities alone cannot meaningfully serve as objectives in the optimisation of active control techniques. The often accepted notion of a decreased $\epsilon$ with successful control \citep{jovanovic-etal-2005, bannier-garnier-sagaut-2016} is shown to be true  only for wall-based control strategies with negligible or no $\power_c$ (passive control). This statement, however, follows from a model assuming that control acts at the wall, and results into the well-known upward shift of the logarithmic portion of the mean velocity profile.

The introduced extended Reynolds decomposition splits the mean velocity profile into a laminar profile $U_\ell$ with the same flow rate, and a profile $U_\Delta$ expressing the zero-integral deviation of the mean profile from $U_\ell$. The turbulent energy production and the dissipation associated to the mean field can thus be expressed as sum of the contributions of the laminar and deviation components, i.e. $\produ = \produ_\ell + \produ_\Delta$ and $\dissU = \dissLam + \dissDelta$, with the single terms being analytically known.
The extended energy box highlights that:
\begin{enumerate}
\item The laminar dissipation $\dissLam$ is the preferable way to dissipate energy, because it is the only dissipation related to a flow rate.
\item The laminar production $\produ_\ell$ corresponds to the fraction of the pumping power wasted by turbulence, and decreases with successful flow control. $\produ_\ell$ may become negative in extreme cases (when sublaminar drag is achieved), and this is seen mathematically since $\produ_\ell$ depends on $\alpha$, Eq.~\eqref{eq:Plam}, and $\alpha$ can become negative, see \eqref{eq:alpha}.
\item The production by deviation $\produ_\Delta$ is the additional power required by a turbulent channel whose mean velocity profile deviates from the laminar one. That any such deviation is detrimental \citep{bewley-2009,fukagata-sugiyama-kasagi-2009} is seen here by the fact that $-\produ_\Delta = \phi_\Delta \ge 0$, i.e. a fraction of $\produ_\ell$ is used to produce mean kinetic energy which is not associated with any flow rate and eventually dissipated by viscosity. However, a reduction of  $\produ_\Delta$ is not a sufficient condition for successful flow control. For instance, $\produ_\Delta$ (and hence $\phi_\Delta$) may be zero also for nonzero $r(y)$ if $3\alpha^2=\beta$ is satisfied, a condition which does not occur for canonical channels and has not been observed for the presently investigated flow control strategies. 
\item While $\produ_\ell$ is the fraction of {\em pumping} power wasted to produce turbulence, $\phi_\Delta + \epsilon$ is the fraction of {\em total} power wasted by turbulence, i.e. the fraction of the total available power not used to produce a flow rate. Therefore $\phi_\Delta + \epsilon$
has to be minimised by control, while $\epsilon$ alone can undergo changes of either sign.
\end{enumerate}

The concepts discussed in the present paper might foster interesting developments. For example the few existing drag-reduction-aware RANS turbulence models \citep[e.g.][]{hassid-poreh-1978,mele-tognaccini-catalano-2016} are often based upon minor modifications of (the model equation for) $\epsilon$ that produce the desired skin-friction reduction effect. The present work has made evident how this approach may lack generality even for passive control. In the flow control domain, knowing the role of the Reynolds stress through the  integrals $\alpha$ and $\beta$ allows a general parametrization of flow control strategies and their energetic effects.

Some limitations of the present work, for instance neglecting control strategies which introduce energy into the mean flow directly, can be remedied easily, while others are more challenging to overcome. Most notably, applying the extended Reynolds decomposition to complex flows is not trivial, as the laminar solution is unknown except for few idealised flow geometries (plane channel, pipe flow, etc.). In case of ducts with arbitrary cross-section, however, the Stokes solution can be adopted instead, as it is readily obtained and is known to yield  minimal dissipation (hence power input) at CFR \citep{fukagata-sugiyama-kasagi-2009}. However, we believe that the significance of the present work mainly resides in its ability to clarify, thanks to both the newly-derived relationships for $\epsilon$ and $\dissU$ and the CPI setting, meaning and objectives of increased energetic efficiency based on skin-friction drag reduction techniques. Much like the FIK identity itself, the limitation of using such tools in the context of simple canonical parallel flows does not diminish their ability to effectively highlight different aspects of the complex physics of near-wall turbulence.

%% file: newboxes-Re.tex
\appendix 
\section{The figure of merit for flow control}
The assessment of flow control strategies in the CPI framework requires the definition of a suitable figure of merit to quantify the control success. A sound figure of merit consistently exceeds a threshold value, known a priori, when flow control is successful. This simple requirement is  satisfied by the ratio $U_b / U_{b,0}$ between the flow rate in the controlled and uncontrolled channel. The control is successful if $U_b / U_{b,0} > 1$, i.e. if flow control yields a higher flow rate than an uncontrolled channel at the same $\power_t$. 

Other possible figures of merit do not fulfil the aforementioned requirement with CPI. Notably, the condition $Re_\tau / Re_{\tau,0}<1$ expressing "less drag" is also verified by unsuccessful active control that wastes power $\power_c$ without positively affecting turbulence, hence just lowering $\power_p$.

Also the drag reduction rate $R$, defined as the relative change of skin-friction coefficient  $C_f=2 \tau_w / \rho U_b^2$, i.e.
\begin{equation}
R \equiv 1 - \frac{C_f}{C_{f,0}} \,,
\label{eq:R-DR}
\end{equation} 
is not a suitable figure of merit for successful control at CPI. In fact, it can be easily shown that $R$ exceeding a certain value implies successful flow control. However, this value is not known a priori and depends upon the unknown value of $\gamma$. This result can be simply obtained by rewriting $C_f$ as $C_f = 2\left(Re_\tau / Re_b \right)^2$ and by inserting the definition \eqref{eq:R-DR} of $R$. One then eliminates $Re_\tau$ in favour of $Re_b$ and $Re_\Pi$ with Eq.~\eqref{eq:CPIconstraint}, to obtain:
\begin{equation}
R = 1 - \left( 1-\gamma \right) \left( \frac{Re_{b,0}}{Re_b} \right)^3 .
\label{eq:R-CPI}
\end{equation}
This expression leads to the conclusion that, at CPI, control is successful (i.e. $Re_b > Re_{b,0}$) only when $R > \gamma$. Thus, the threshold value for $R$ is control-dependent. 

Among the other possible figures of merit, we also mention drag reduction rate, net power saving and gain defined according to \cite{kasagi-hasegawa-fukagata-2009}. These quantities are meaningful only for flows compared at CFR: simple algebraic steps like those described above show why they do not work at CPG or CPI.

\section{$Re$ dependence of the energy fluxes in the uncontrolled channel flow}

\begin{figure}
\centering
\includegraphics[]{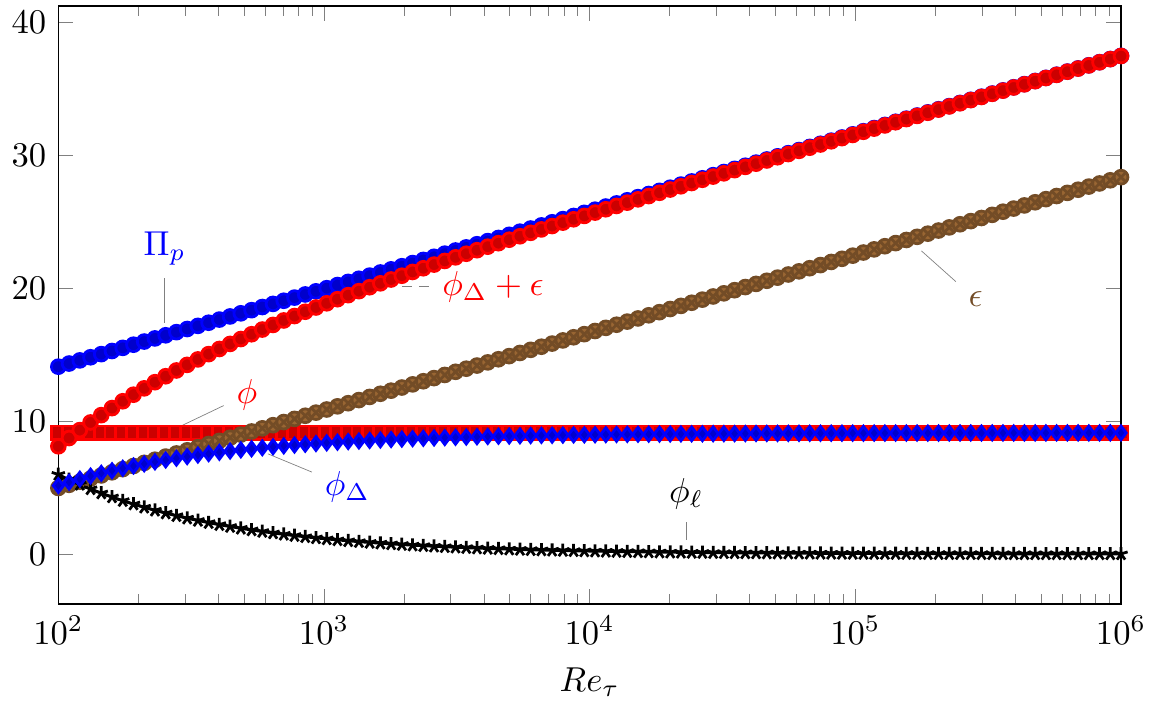}
\caption{Classic and present decomposition of the energy fluxes appearing in the extended energy box, expressed in viscous units as a function of $Re_\tau$.}
\label{fig:abe-retau}
\end{figure}

A description of how the various energy fluxes in the extended energy box change their relative importance with $Re$ can be obtained by resorting to empirical relationships like those discussed by \cite{abe-antonia-2016}. Figure \ref{fig:abe-retau} represents all the energy fluxes, in viscous units, and shows how they change with $Re_\tau$. The empirical fits $\dissU^+ = 9.13$ and $\epsilon^+ = 2.54 \log Re_\tau - 6.72$ are used \citep{abe-antonia-2016}. The usual decomposition shows that $\dissU^+$ is constant and represents the largest contribution to viscous dissipation at low $Re$, while $\epsilon^+$ grows logarithmically, becomes equal to $\dissU^+$ at $Re_\tau \approx 600$, and eventually accounts for the majority of dissipation. On the other hand, $\dissLam^+$ is important at low $Re$ only, and eventually disappears for $Re_\tau > 1000$, where $\dissDelta^+$ becomes nearly coincident with $\dissU^+$.

\begin{figure}
\centering
\includegraphics[]{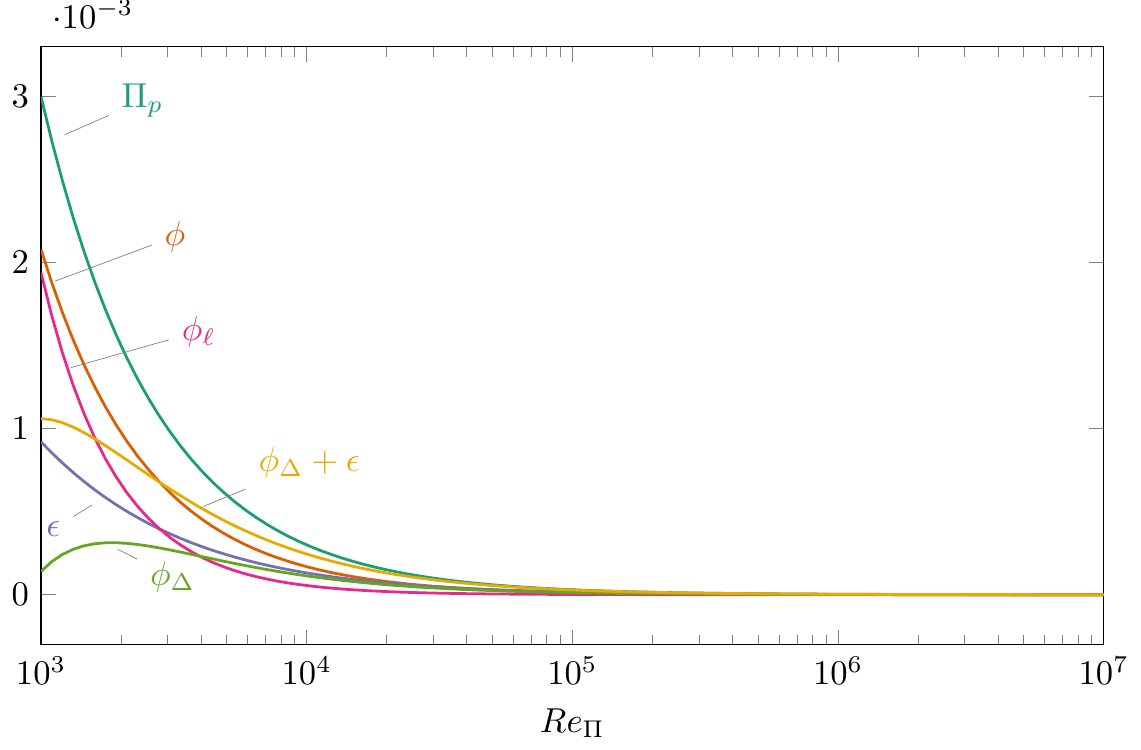}
\caption{Classic and present decomposition of the energy fluxes appearing in the extended energy box, expressed in power units as a function of $\Reyn$.}
\label{fig:abe-repi}
\end{figure}

\begin{figure}
\centering
\includegraphics[]{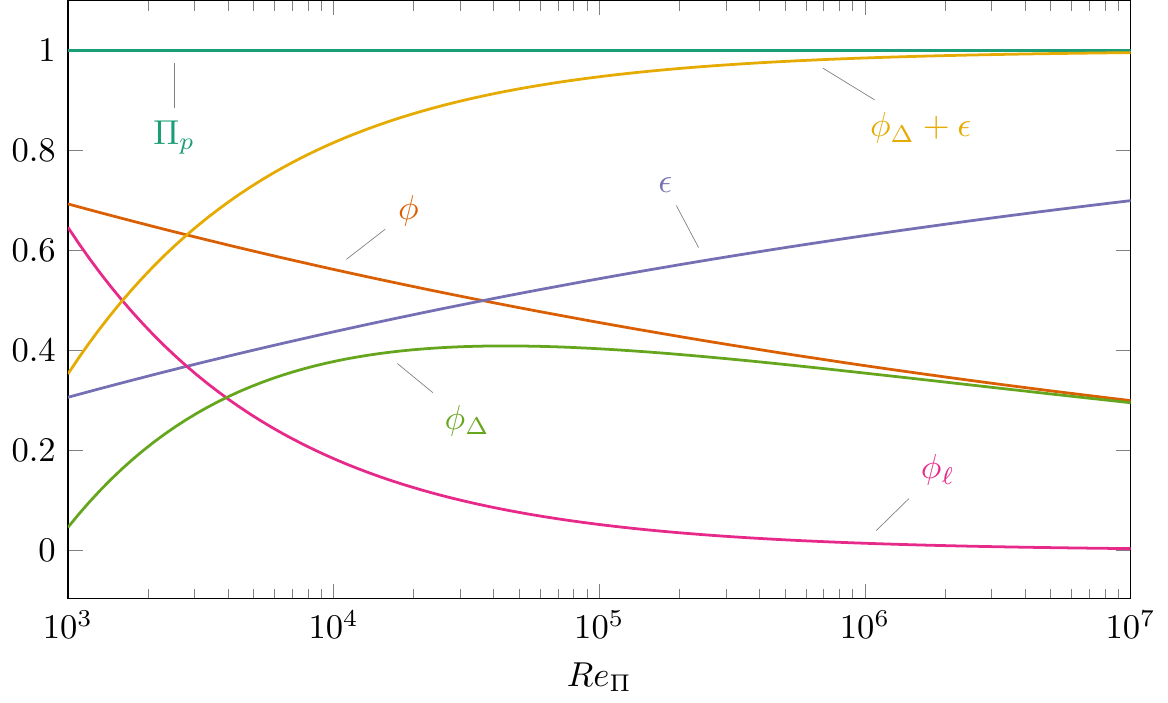}
\caption{Classic and present decomposition of the energy fluxes appearing in the extended energy box, expressed in power units as a function of $\Reyn$ and normalized with $\power_p$.}
\label{fig:abe-repi-normalized}
\end{figure}
Figure \ref{fig:abe-repi} plots the same quantities but with power-based nondimensionalization. In this representation, all quantities tend to zero for $Re \rightarrow \infty$, as shown by Eq.\eqref{eq:total-power}. Figure \ref{fig:abe-repi-normalized} is a replot of Fig.~\ref{fig:abe-repi}, but every quantity is now normalised by $\power_p$. It is interesting to observe how the cross-over point where $\dissU = \epsilon$, known to occur at $Re_\tau \approx 600$ or $\Reyn \approx 36000$, corresponds in this graph to the maximum of $\dissDelta / \power_p$. Moreover, the quantity $\dissDelta + \epsilon$ is observed to equal the pumping power in the limit of $Re \rightarrow \infty$. We note that  $\dissDelta + \epsilon = \produ_\ell$ is ``wasted'' total power which does not produce any flow rate, and therefore quantifies the potential increase in flow rate if relaminarization of the flow is achieved. At CPI, this characterisation of the high-$Re$ limit allows us to identify the asymptotic behaviour of expressions like \eqref{eq:Ub-alpha}, \eqref{eq:phi_lam} and \eqref{eq:Plam}. This is indeed not trivial, as there the group $(\alpha \Reyn)$ appears, and $\alpha \rightarrow 0$ when $\Reyn \rightarrow \infty$.

It is first shown that $\alpha \rightarrow 0$ for infinite $\Reyn$. A preliminary step is to realise that the bulk velocity normalised with the power based velocity tends to zero at infinite Reynolds number; i.e. $U_b \rightarrow 0$ for $\Reyn \rightarrow \infty$. Based on Eq.~\eqref{eq:FIK_U_2} and  Eq.~\eqref{eq:power-p} one finds
\begin{equation}
1 = \lim_{\Reyn \rightarrow \infty} \frac{\produ_\ell}{\power_p} = \frac{U_b \alpha \Reyn}{1-\gamma} .
\label{eq:limit-Pl}
\end{equation}
The FIK relation \eqref{eq:FIK_U_3} gives that $U_b \alpha \Reyn = 1 - \gamma - U_b^2$, and substituting into the above \eqref{eq:limit-Pl} leads to:
\[
1 = \lim_{\Reyn \rightarrow \infty} \frac{\produ_\ell}{\power_p} = 1 - \frac{U_b^2}{1-\gamma}
\]
implying that $U_b \rightarrow 0$.

Now, the asymptotic behaviour for $\alpha$ can be easily arrived at by introducing the limit $U_b \rightarrow 0$ for $\Reyn \rightarrow \infty$ into Eq. \eqref{eq:FIK_U_1}, which implies that $\alpha$ must tend to zero. If this is the case, the asymptotic behaviour of \eqref{eq:limit-Pl} requires $\alpha \Reyn \rightarrow \infty$. This is the correct limit to take, even though $\alpha \rightarrow 0$.

In the limit of $\alpha \Reyn \rightarrow \infty$, Eq.~\eqref{eq:Ub-alpha} confirms that $U_b \rightarrow 0$; Eq.~\eqref{eq:phi_lam} shows that $\dissLam \rightarrow 0$, which agrees with the visual observation in Fig.~\ref{fig:abe-repi-normalized}, and, consistently, from Eq.~\eqref{eq:Plam} it can be seen that $\produ_\ell \rightarrow 3 (1-\gamma)/\Reyn = \power_p$, indicating that the entire pumping power will be transferred to turbulence.

The functional form for $\dissU^+$ and $\epsilon^+$, discussed in the context of Fig.~\ref{fig:ebox-extended-nocontrol}, implies a specific relationship between $Re_b$ and $Re_\tau$. Nondimensionalized in power units, the viscous dissipation $\dissU + \epsilon$ is an explicit function of $\Reyn$, given by Eq.\eqref{eq:power-p}, i.e. $\dissU + \epsilon = 3/\Reyn$. Combining this expression with the empirical fits by \cite{abe-antonia-2016} for $\dissU$ and $\epsilon$ recast in power units, one obtains the following relationship between $\Reyn$ and $Re_\tau$:
\begin{equation*}
\Reyn = \left( 0.803 + 0.85 \log Re_\tau \right)^{0.5} Re_\tau^{3/2} .
\end{equation*}
Eliminating $\Reyn$ with Eq.~\eqref{eq:CPIconstraint} from the above formula, one obtains for $\gamma=0$:
\begin{equation}
Re_b = \left( 2.41 + 2.55 \log Re_\tau \right) Re_\tau .
\label{eq:ReB-ReTau}
\end{equation}

This formula is analogous to the one derived by \cite{hasegawa-quadrio-frohnapfel-2014} based on the Deans' correlation \citep{dean-1978} only, which reads
\begin{equation}
Re_b = 7.3204 Re_\tau^{8/7} .
\label{eq:ReB-ReTau-Dean}
\end{equation}

By comparing the two expressions \eqref{eq:ReB-ReTau} and \eqref{eq:ReB-ReTau-Dean}, one notices the slightly different $Re_\tau$ exponents, and -- most importantly -- the weak, logarithmic dependence of the prefactor upon $Re_\tau$ in Eq.~\eqref{eq:ReB-ReTau}. Regardless of the numerical values of the coefficients, the appearance of $Re_\tau$ in the prefactor is a consequence of $\epsilon^+$ being a function of $Re_\tau$. Hence, the Deans' correlation in the form \eqref{eq:ReB-ReTau-Dean} implies that $\epsilon^+$ does not vary with $Re_\tau$. A similar observation holds for other empirical relationships linking $Re_b$ and $Re_\tau$ obtained by least-square fitting of skin-friction measurements at different values of $Re$ and presented in the form $C_f= 2\left( Re_\tau / Re_b \right)^2 = f\left( Re_b \right)$ \cite[see for instance][]{dean-1978, zanoun-nagib-durst-2009, schultz-flack-2013}.